\documentclass[showpacs,showkeys,aps,prd,floatfix]{revtex4}
\usepackage{amssymb}
\usepackage{amsmath}
\usepackage{graphicx}
\usepackage{bm}
\usepackage{bbm}
\usepackage{tensor}
\usepackage{slashed}
\usepackage{tikz,ifthen}
\usepackage{tikz}
\usepackage{tikz,ifthen}

\usetikzlibrary{trees}
\usetikzlibrary{decorations.pathmorphing}
\usetikzlibrary{decorations.markings}
\tikzset{beamerprimary/.style={structure.fg, thick}}
\tikzset{beamersecondary/.style={structure.bg, thick}}
\tikzset{ boson/.style={decorate, decoration={snake}},
     gauge/.style={decorate,decoration={snake,post length=1mm}}  ,
     fermion/.style={postaction={decorate},
        decoration={markings,mark=at position .55 with {\arrow{>}}}},
    fermionloop/.style={postaction={decorate},
        decoration={markings,mark=at position .25 with {\arrow{<}}}}, 
    gluon/.style={decorate, 
        decoration={coil,amplitude=4pt, segment length=5pt}},
    scalar/.style={dashed},
    graviton/.style={double},
    tdm/.style={double,postaction={decorate},decoration={markings,mark=at position .55 with {\arrow{>}}}},
    tdmloop/.style={double,postaction={decorate},
        decoration={markings,mark=at position .25 with {\arrow{<}}}} 
}

\begin{document}

\title{Space-time origin of gauge symmetry}
\author{ Mauro Napsuciale}
\affiliation{Departamento de F\'{\i}sica, Universidad de Guanajuato, Lomas del Bosque
103, Fraccionamiento Lomas del Campestre, C. P. 37150, Le\'{o}n, Guanajuato,
M\'{e}xico.}

\begin{abstract}
In this work I show by a first principles calculation that quantum states describing massive relativistic free spinning 
particles obey kinematical conditions whose origin can be traced to parity as a good quantum number. These conditions 
are at the root of the equations of motion and of the constraints satisfied by the corresponding fields. In the massless 
limit, well defined parity is lost but a symmetry emerges related to arbitrary changes in the unphysical parity components.
It is shown that this emergent symmetry is the celebrated gauge symmetry.
\end{abstract}

\keywords{Dark matter}
\maketitle

\section{Introduction}
The standard model is a quantum theory of elementary particle interactions structured in terms of three basic elements: i) the equations 
of motion for free particles with spin $j=0,1/2,1$; ii) the gauge principle and iii) the mechanism of spontaneous symmetry breaking. 
It was devised starting from the electromagnetic interactions for charged particles where the classical Lorentz force, raised to a minimal 
substitution principle at the Hamiltonian classical level, is used in the early formulations of quantum mechanics 
\cite{Weyl:1928wq,Weyl:1931wh} (for a historical account of the gauge principle see \cite{Jackson:2001ia}). 
The gauge principle connects three previously disconnected concepts belonging to different realms: the gauge invariance 
of the classical electromagnetic field, the Lorentz force and the global phase invariance of quantum mechanics. After the formulation 
of the Klein-Gordon \cite{Fock:1926fj, Klein:1926tv, Gordon:1926ov}, Dirac \cite{Dirac:1930ek} and Proca \cite{Proca:1936fbw} 
equations of motion for relativistic particles, the quantization reveals these as quantum theories for systems 
with many particles and anti-particles. In the absence of interactions, in addition to the ground state (the vacuum), states are 
characterized with two conserved natural numbers which define the contained number of free particles and antiparticles respectively. 
Interactions induce transitions between these sectors with well defined number of particles giving rise to electromagnetic scattering processes. 

Obtaining a gauge theory for the weak interactions of leptons required to go from the effective Fermi interaction \cite{Fermi:1934hr}
to the intermediate massive vector boson hypotesis \cite{Lee:1949qk}, the discovery of parity violation \cite{Lee:1956qn,Wu:1957my} 
and the V-A structure \cite{Sudarshan:1958vf,Feynman:1958ty} to the formulation of non-Abelian gauge symmetries \cite{Yang:1954ek} 
and the discovery of alternative mechanisms in nature for the implementation of symmetries at the quantum level (the spontaneous 
breaking of symmetries \cite{Nambu:1960tm,Nambu:1961tp,Nambu:1961fr,Goldstone:1961eq,Goldstone:1962es,Englert:1964et,Higgs:1964ia,Higgs:1964pj,Guralnik:1964eu}) 
such that massive weak gauge fields can be understood as massless fields in a broken phase \cite{Weinberg:1967tq,Salam:1968rm}. 
In the hadron sector it was only after the classification of hadrons in $SU(3)_{F}$ higher multiplets
 \cite{Gell-Mann:1961omu,Neeman:1961jhl} and the speculation on the possibility that also the fundamental representations have 
realization in nature \cite{Zweig:1964ruk,Gell-Mann:1964ewy}, that the strong interactions could be understood at the quark level 
as an $SU(3)_{c}$ gauge theory \cite{Greenberg:1964pe, Han:1965pf,Fritzsch:1973pi}. The renormalization of massive Yang-Mills 
theories stablished in \cite{tHooft:1971qjg} and the calculation of the beta function for non-Abelian gauge theories exhibiting ultraviolet 
asymptotic freedom \cite{Gross:1973id,Politzer:1973fx} yield a gauge theory for strong and electroweak interactions with an 
$SU(3)_{c}\otimes SU(2)_{L} \otimes U(1)_{Y}$ gauge group. The consistency of this formalism at the quantum level, requires the 
cancellation of the effects breaking the standard model symmetries by quantum 
corrections (anomalies) \cite{Adler:1969gk, Bell:1969ts, Delbourgo:1972xb, Witten:1982fp, Geng:1989tcu}. 

The amazing summary of more than a century of particle physics is that electromagnetic, weak and strong interactions of elementary 
particles are described by a quantum field theory formalism with a well defined particle content (three families of massless fermions) a 
gauge principle (introducing massless gauge fields) and a spontaneous breaking 
$SU(3)_{c}\otimes SU(2)_{L} \otimes U(1)_{Y}\to SU(3)_{c}\otimes U(1)_{em}$ which yields mass to fermions and weak gauge bosons.
In spite of the wonder of this result, emergent phenomena as dark matter or dark energy, ancient unsolved 
mysteries as the matter-antimatter asymmetry and the long standing problem of the mass and nature of neutrinos, requires to go 
beyond the standard model and it was expected that particles proposed in a variety of possibilities for physics beyond the standard 
model show up at the $TeV$ scale, but experimental data from the Large Hadron Collider yields no trace of new particles so far.  

Sometimes, a different  look at our description of nature yields new understanding of the corresponding phenomena and new 
perspectives on the unsolved problems. In this concern, it must be said that the gauge principle works, but we do not have a deep 
understanding of it, except that in the case of electromagnetism it is just a rewriting of the Lorentz force. On the other hand, the equations 
of motion for standard model free particles have been formulated trying to solve specific problems at the time ( negative probabilities of the 
Klein-Gordon equation in the case of Dirac \cite{Dirac:1930ek} and to establish a theory which exclude the negative energy solutions of 
Dirac equation in the case of Proca \cite{Proca:1936fbw}), but the conceptual systematics behind these equations is not clear, a problem
related to the ancient desire of having a consistent quantum description of higher spin particles
\cite{Fierz:1939ix,Bargmann:1948ck,Bhabha:1945zz,Harish-Chandra:1947okk,Hurley:1971ipy,Tung:1967zz,Niederle:2001wgk,
Hurley:1974xj,Good:1989ja}. 

In this paper, by a first principles calculation I show that the equations of motion for massive free particles have their origin in the 
discrete symmetry of parity. There have been attempts in the literature to 
deduce general principles behind the equations of motion for free relativistic particles where parity plays a secondary role. In fact, the 
Dirac equation has been derived from a calculation of the fundamental representations of the proper orthochronous homogeneous 
Lorentz group in the Ryder's book \cite{Ryder:1985wq} which are connected by parity and a principle of indistinguishability. 
As for the Proca equation, an {\it{ab initio}} construction of states in the 
$(1/2,1/2)$ representation considering $(1/2,1/2)=(1/2,0)\otimes (0,1/2)$ has been done in \cite{Ahluwalia:2000pj} focusing first in the  
states of well defined angular momentum and constructing later a Lorentz invariant inner product where parity appears naturally. 
The parity eigen-subspaces coincide with the subspaces of well defined angular momentum in this representation which allows to 
write the parity condition in a frame where the particle has momentum $\bm{p}$ in the form of a second order matrix equation where the 
state is represented by a column vector with four components. In \cite{Napsuciale:2002ny} it was shown that this equation can be derived 
directly from the construction of parity from the fact that it maps $(1/2,0)\otimes (0,1/2)$ into the unitarily equivalent $(0,1/2)\otimes (1/2,0)$ 
representation and using the Ryder's principle of indistinguishability. 

Here I will do a first principles derivation of the Dirac and Proca equations. By first principles I mean simply what we have learnt 
from quantum mechanics during the last century and we can summarize in six axioms which lead us directly to the conclusion that 
the complete set of commuting observables (CSCO) whose eigenvalues (the good quantum numbers) characterize a quantum 
system, can be obtained from its full group of symmetries. A free relativistic massive elementary particle is a quantum system and 
its full group of symmetries is the set of transformations leaving invariant the space-time length $ds^{2}=dx^{\mu}dx_{\mu}$, the 
Poincar\'{e} group. This is of course well known but, amazingly, the conventional analysis consider only the proper isochronous 
subgroup of the Poincar\'{e} group leaving aside the crucial role of discrete symmetries in the characterization of the quantum 
states for a free particle. I show that parity must be in the CSCO and this impose kinematical conditions on the  free particle states 
which are nothing else that the known equations of motion. Furthermore, an explicit construction of the quantum states 
of defined Poincar\'{e} quantum numbers in the $(1/2,1/2)$ representation of the proper orthochronous homogeneous Lorentz 
group shows that, 
in addition to the kinematical condition, a constraint is obtained which eliminates the unphysical parity component of a general state 
residing in this representation. In the massless limit this constraint is lost but a symmetry emerges related to arbitrary changes 
in the unphysical parity sector an we find that, in a particular basis for the $(1/2,1/2)$ space, this emergent symmetry is the 
gauge symmetry of a massless vector field. 

A similar structure is obtained for the $(1,0)\oplus (0,1)$ representation space. Although the standard model does not contain 
dynamical fields transforming in this representation, it has been pointed out that its algebraic structure has appealing 
properties for the description of dark matter and this possibility has been recently explored 
\cite{Gomez-Avila:2013qaa,Napsuciale:2015kua,Hernandez-Arellano:2018sen, Hernandez-Arellano:2019qgd}.
The leading terms in the effective theory for $(1,0)\oplus (0,1)$ dark matter interactions with standard model fields turn out to have canonical 
dimension four, which opens the door for the construction of a renormalizable theory. However, the 
structure of the propagator of these fields clearly yield divergent contributions at high energies, similarly to the intermediate vector 
bosons for the Fermi effective theory of weak interactions \cite{Lee:1949qk}. The present work is motivated by the need to clarify if a proper 
behavior at high energies of the $(1,0)\oplus (0,1)$ propagator can be constructed using mass generation for these fields by alternative 
mechanisms analogous to the Higgs mechanism in the standard model. A first step in this direction is to study the properties of 
the massless limit which drove me to the connection between parity and the gauge symmetry. 

The paper is organized as follows. In the next section I clearly establish what I mean by first principles, construct the complete set of 
commuting observables for a free massive elementary particle and establish the kinematical condition imposed by parity for the 
corresponding quantum states. In Section III I review the calculation of the irreducible representations of the proper isochronous 
homogeneous Lorentz group, solve its algebra for the $(j,0)$ and $(0,j)$ representations and show that the kinematical condition 
imposed by parity yields the Dirac equation for  $j=1/2$ and a spinor-like equation for the case  $j=1$. Section IV is devoted to 
the first principles calculation of the {\it quantum} $(1/2,1/2)$ representation including the massless limit. In Section V 
I address the massless limit of the $(1,0)\oplus (0,1)$ representation and we close with our conclusions in Section VI.
 
\section{First principles, parity and kinematical condition}
During the past century we have learnt that every quantum system is 
characterized by its symmetries. Indeed, we can summarize our understanding of quantum systems and measurement of observables 
with classical apparatus in what is called axiomatic quantum mechanics which translate results of experiments with quantum systems 
into the following axioms:
\begin{enumerate}
\item All the information of a physical system is encoded in a vector state $|\psi (t)\rangle$ depending parametrically on the time $t$ and 
living in a Hilbert space ${\cal{H}}$.
\item Observables are represented by Hermitian operators in ${\cal{H}}$.
\item The measurement of an observable $A$  can yield as a result only one of the eigenvalues $a_{i}$ of the corresponding Hermitian
operator ${\cal{A}}$.
\item Before the measurement we do not know which one of the eigenvalues will result, we only know the probability, which is given by
\begin{equation}
{\cal{P}}(a_{k},t)=|\langle a_{k}|\psi (t)\rangle|^{2}.
\end{equation}
\item The state of the system after the measurement is given by the eigenstate $|a_{m}\rangle$ corresponding to the resulting eigenvalue 
$a_{m}$.
\item The evolution in time of the state $|\psi (t)\rangle$ is generated by the Hamiltonian of the system
\begin{equation}
i\hbar\partial_{t} |\psi (t)\rangle = H (p,q,t)|\psi (t)\rangle.
\end{equation}
\end{enumerate}

Eigenstates of Hermitian operators belonging to different eigenvalues are orthogonal and this property can be used to get a 
complete characterization of the Hilbert space. Starting with any observable we perform a first characterization which is 
refined if we have degenerate 
eigenvalues successively incorporating new commuting observables which lands in a complete set of commuting 
observables (CSCO) whose common eigenstates form a basis for the Hilbert space and are labelled by the corresponding eigenvalues 
called good quantum numbers. This procedure is democratic with respect to the choice of the observables but the Hamiltonian 
has a preferred role because it dictates the dynamics and we include it in the first place, in such a way that all the remaining operators 
in the CSCO commute with $H$ and among themselves. Here is where symmetries enter into scene because symmetry transformations
must be implemented by linear (or anti-linear) unitary operators \cite{Wigner:1931cj}. After $H$, the natural candidates are the Casimir 
operators of the full group of symmetries of the quantum system. The full group of symmetries may contain a Lie subgroup of continuous 
symmetries connected to the identity operator by the exponential map, and we can write the symmetry operators in terms of continuous 
parameters and a set of generators (one for each parameter) which turn out to be Hermitian operators. The generators necessarily 
commute with the Hamiltonian and  are natural candidates for the CSCO. However, only those commuting with themselves (those 
which generate the Cartan subalgebra) can belong to this set. Finally, we can have also discrete symmetries in the full group of 
symmetries of the quantum system. If the corresponding operators are also Hermitian they are also candidates to be in the CSCO. 

In summary, the symmetries of a quantum system yield a complete characterization of the quantum space of states in terms of the 
eigenvalues of the Hamiltonian, the Casimir operators of the full group of symmetries, the generators of the Cartan subalgebra and 
the Hermitian discrete symmetries operators which commute with these operators.

An elementary particle is a quantum system and a free elementary particle has the same symmetries as the Minkowski space-time. 
This lead us to the classical transformations leaving invariant the length element $ds^{2}=dx^{\mu}dx_{\mu}$ 
and the later implementation of these transformations in the Hilbert space of quantum mechanics. The most general transformation 
leaving invariant $ds^{2}$ is of the form
\begin{equation}
x^{\prime\mu}=\Lambda^{\mu}_{~\nu}x^{\nu}+a^{\mu} \equiv T^{\mu}_{~\nu}(\Lambda, a)x^{\nu}.
\end{equation}
The identity transformation is $T(\mathbbm{1},0)$ and the composition of two Poincar\'{e} transformations yields
\begin{equation}
T(\Lambda_{2}, a_{2})T(\Lambda_{1}, a_{1}) =  T(\Lambda_{2} \Lambda_{1}, \Lambda_{2} a_{1} +  a_{2} ).
\end {equation}
It can be shown that this composition is associative and the inverse is given by
\begin{equation}
T^{-1}(\Lambda,a)=T(\Lambda^{-1},-\Lambda^{-1}a),
\end{equation}
thus Poincar\'{e} transformations form a group. From the invariance of the length element we obtain
\begin{equation}
\Lambda^{t}g\Lambda=g,
\label{inv}
\end{equation}
where $(g)_{\mu\nu}$ is the metric matrix $g=Diag (1,-1-1-1)$, thus 
\begin{equation}
\Lambda^{-1}=g^{-1} \Lambda^{t} g.
\end{equation}
From the composition rule we can see that every Poincar\'{e} transformation 
can be written as the product of a space-time translation and an homogeneous Lorentz transformation
\begin{equation}
 T(\Lambda, a)= T(\mathbbm{1},a) T(\Lambda,0).
\end{equation}
Space-time translations form an invariant Abelian subgroup
\begin{align}
T(\mathbbm{1},a_{2})T(\mathbbm{1},a_{1})&=T(\mathbbm{1},a_{1}+a_{2})=T(\mathbbm{1},a_{1})T(\mathbbm{1},a_{2}), \\
 T(\Lambda,b) T(\mathbbm{1},a)T^{-1}(\Lambda, b) &= T(\mathbbm{1},\Lambda a ).
\end{align}
Homogeneous transformations $T(\Lambda,0)$ form also a subgroup, named Homogenous Lorentz Group (HLG), with the 
composition rule
\begin{equation}
T(\Lambda_{2},0)T(\Lambda_{1},0)=T(\Lambda_{2}\Lambda_{1},0).
\end{equation}  
From Eq.(\ref{inv}) we obtain $\det \Lambda= \pm 1$ and 
from the $00$ component of  Eq.(\ref{inv})  we get $(\Lambda^{0}_{~0})^{2}\geq 1$. It can be shown that if we perform two 
succesive homogeneous Lorentz 
transformation $\bar\Lambda\Lambda$ the sign of $\Lambda^{0}_{~0}$ is the same as the sign of $(\bar\Lambda\Lambda)^{0}_{~0}$,
thus we get a classification in terms of four disconnected subsets 
characterized by the determinant of the matrix $\Lambda$ and the sign of  $\Lambda^{0}_{~0}$. The proper ($det(\Lambda)=1$) 
orthochronous ($\Lambda^{0}_{~0}\geq 1$) transformations contains the identity transformation and form a subgroup, the subgroup 
of proper isochronous homogeneous Lorentz transformations, $L^{\uparrow}_{+}$, composed of rotations and changes of inertial 
frame (boosts). 

Transformations in other subsets can be obtained from the composition of the discrete transformations, parity, time reversal 
and their composition, with transformations in $L^{\uparrow}_{+}$. In the Minkowski space these transformations are the mappings 
${\cal{P}}: (x^{0},x^{i})\to (x^{0},-x^{i})$,  ${\cal{T}}: (x^{0},x^{i})\to (-x^{0},x^{i})$ and 
${\cal{R}_{T}}\equiv{\cal{PT}}: (x^{0},x^{i})\to (-x^{0},-x^{i})$ respectively. 
Clearly, these transformations leave invariant the length element in Minkowski space and as classical Poincar\'e transformations have 
the matrix representation 
\begin{equation}
(T({\cal{P}},0))^{\mu}_{~\nu}=Diag (1,-1-1-1), \quad (T({\cal{T}},0))^{\mu}_{~\nu}= Diag (-1,1,1,1), 
\quad (T({\cal{R}_{T}},0))^{\mu}_{~\nu}= Diag (-1,-1,-1,-1).
\end{equation}
This paper focus on the crucial role of discrete symmetries at the quantum level and it is important to asses its action at the 
classical level. For ${\cal{D}}={\cal{P}}, {\cal{T}}, {\cal{R}_{T}}$, the composition rule of the Poincar\'{e} group yields
\begin{equation}
T({\cal{D}},0)T(\Lambda, a)T^{-1}({\cal{D}},0)=T({\cal{D}}\Lambda {\cal{D}}^{-1},{\cal{D}}a).
\end{equation}
Also important below will be the action of a Poincar\'{e} transformation on the discrete transformations i.e.
\begin{equation}
T(\Lambda, a)T({\cal{D}},0)T^{-1}(\Lambda, a)=T(\Lambda{\cal{D}},a) T(\Lambda^{-1},-\Lambda^{-1} a) 
= T(\Lambda{\cal{D}}\Lambda^{-1},-\Lambda{\cal{D}}\Lambda^{-1} a + a).
\end{equation}
 
At the quantum level we require to work out the action of the Poicar\'e transformations on Hilbert space. These are represented by 
linear (or anti-linear) unitary operators
\begin{equation}
|\psi\rangle \rightarrow U(\Lambda, a)|\psi\rangle,
\end{equation}
satisfying the composition rule
\begin{equation}
U(\Lambda_{2},a_{2})U(\Lambda_{1},a_{1})=U(\Lambda_{2} \Lambda_{1}, \Lambda_{2}a_{1}+ a_{2}).
\end{equation}
This rule yields the following general results
\begin{align}
U(\Lambda,a)&=U(\mathbbm{1},a)U(\Lambda,0) \label{factorization} \\
U^{-1}(\Lambda,a)&=U(\Lambda^{-1}, -\Lambda^{-1} a) \\
U(\mathbbm{1},a_{2})U(\mathbbm{1},a_{1})&=U(\mathbbm{1},a_{1}+a_{2}), \\
U(\Lambda,b) U(\mathbbm{1},a)U^{-1}(\Lambda,b) &=U(\mathbbm{1},\Lambda a )
\end{align}
Let us consider first the continuous Poincar\'{e} transformations connected to the identity, $U(a,\Lambda)$ with $\Lambda\in L^{\uparrow}_{+}$, 
and write them in terms of the generators
\begin{equation}
U(a, \Lambda)=e^{-i P^{\mu}a_{\mu} - \frac{i}{2}J^{\mu\nu}\theta_{\mu\nu} },
\end{equation}
where $J^{\mu\nu}$ is an antisymmetric tensor containing the generators of rotations, $\bm{J}=(J^{23},J^{31},J^{12})$ and boosts 
$\bm{K}=(J^{01},J^{02},J^{03})$, and $P^{\mu}$ are the generators of space-time translations. Taking infinitesimal transformations it can be shown that the composition rule of the group impose the the following algebra for the generators
\begin{align}
[P^{\mu},P^{\nu}]&=0, \label{comrelPP} \\
[J^{\mu\nu},P^{\alpha}]&=i(g^{\mu\alpha}P^{\nu}-g^{\nu\alpha}P^{\mu}) \label{comrelJP} \\
[J^{\mu\nu},J^{\alpha\beta}]&=i(g^{\mu\alpha}J^{\nu\beta}-g^{\nu\alpha}J^{\mu\beta}
-g^{\mu\beta}J^{\nu\alpha}+g^{\nu\beta}J^{\mu\alpha}) .
\label{comrelJJ}
\end{align}

Let us consider now discrete transformations. 
These are homogeneous transformations and must follow the composition rule of the 
group, thus there must be  operators $U({\cal{D}}, 0)$ such that
\begin{equation}
U({\cal{D}}, 0)U(\Lambda, a)U^{-1}({\cal{D}}, 0)=U({\cal{D}}\Lambda {\cal{D}}^{-1}, {\cal{D}}a ).
\label{parpoin}
\end{equation}
Considering a $U(\Lambda, a)$ in the proper orthochronous subgroup and infinitesimal transformations we arrive at
\begin{align}
U({\cal{D}}, 0)(i P^{\mu})U^{-1}({\cal{D}}, 0)&= i {\cal{D}}^{~\mu}_{\alpha} P^{\alpha} 
\label{dispropP} \\
U({\cal{D}}, 0)(i J^{\mu\nu})U^{-1}({\cal{D}}, 0)&=i {\cal{D}}^{~\mu}_{\alpha} {\cal{D}}^{~\nu}_{\beta} J^{\alpha\beta}.
\label{dispropJ}
\end{align}
We conserved the $i$ factors in these relations in order to include the possibility of  anti-linear unitary (anti-unitary) 
realization for these transformations pointed by Wigner long ago \cite{Wigner:1931cj}. In Particular for $\mu=0$ in Eq.(\ref{dispropP}) 
we get
\begin{align}
U({\cal{P}}, 0)(i H)U^{-1}({\cal{P}}, 0)&=  i  H, \label{PpropiH} \\
U({\cal{T}}, 0)(i H)U^{-1}({\cal{T}}, 0)&= - i  H, \label{TpropiH} \\
U({\cal{R}_{T}}, 0)(i H)U^{-1}({\cal{R}_{T}}, 0)&=  -i  H.  \label{RpropiH}
\end{align}
The minus sign on the right hand side of Eqs. (\ref{TpropiH},\ref{RpropiH}) requires $U({\cal{T}}, 0)$ and $U({\cal{R}_{T}}, 0)$ 
to be realized as anti-unitary operators in Hilbert space. Indeed, a linear realization would require that for every
eigenstate of $H$ 
\begin{equation}
H | E\rangle = E | E\rangle
\end{equation}
the transformed state under $U({\cal{T}}, 0)$ and $U({\cal{R}_{T}}, 0)$ must satisfy
\begin{equation}
H (U({\cal{D}}, 0) | E\rangle) = - E (U({\cal{D}}, 0)| E\rangle) , \qquad {\cal{D}}={\cal{T}},{\cal{R}_{T}},
\end{equation}
yielding free particles with negative energy. Anti-linearity cancel this minus sign yielding degeneracy instead of 
negative energies, thus $U({\cal{P}}, 0)$ can be realizad as a linear unitary operator but $U({\cal{T}}, 0)$ and 
$U({\cal{R}_{T}}, 0)$ must have an anti-linear unitary realization. With these considerations the specific transformation properties 
under discrete symmetries for the components of $P^{\mu}$ and $J^{\mu\nu}$ are
\begin{align}
U({\cal{P}}, 0)HU^{-1}({\cal{P}}, 0)&=  H, \qquad U({\cal{P}}, 0)\bm{P}U^{-1}({\cal{P}}, 0)= -\bm{P} \label{PpropP} , \\
U({\cal{T}}, 0)HU^{-1}({\cal{T}}, 0)&=   H, \qquad U({\cal{T}}, 0)\bm{P}U^{-1}({\cal{T}}, 0)= -\bm{P}\label{TpropP} ,\\
U({\cal{R}_{T}}, 0) HU^{-1}({\cal{R}_{T}}, 0)&=   H, \qquad U({\cal{R}_{T}}, 0) \bm{P}U^{-1}({\cal{R}_{T}}, 0)=   \bm{P},  \label{RpropP} \\
U({\cal{P}}, 0)\bm{J}U^{-1}({\cal{P}}, 0)&=  \bm{J}, \qquad U({\cal{P}}, 0)\bm{K}U^{-1}({\cal{P}}, 0)= -\bm{K} \label{PpropJK} , \\
U({\cal{T}}, 0)\bm{J}U^{-1}({\cal{T}}, 0)&=  - \bm{J}, \qquad U({\cal{T}}, 0)\bm{K}U^{-1}({\cal{T}}, 0)= \bm{K}\label{TpropJK} ,\\
U({\cal{R}_{T}}, 0) \bm{J}U^{-1}({\cal{R}_{T}}, 0)&= - \bm{J}, \qquad U({\cal{R}_{T}}, 0) \bm{K}U^{-1}({\cal{R}_{T}}, 0)= -  \bm{K}.  \label{RpropJK}
\end{align}

Now we need to identify the Casimir operators of the group in order to choose a CSCO for free particle states.
The first Casimir operator of the proper orthochronous Poincar\'{e} subgroup is $P^{\mu}P_{\mu}$. The second Casimir operator 
is the square of the Pauli-Lubanski operator $W_{\mu}=\frac{1}{2}\epsilon_{\mu\alpha\beta\rho}J^{\alpha\beta}P^{\rho}$. This operator satisfy $P^{\mu}W_{\mu}=0$ and the following 
commutation relations 
\begin{align}
[W_{\mu},P_{\nu}]&=0, \\
[J^{\mu\nu},W^{\alpha}]&=i(g^{\mu\alpha}W^{\nu}-g^{\nu\alpha}W^{\mu}), \\
[W_{\mu},W_{\nu}]&= i\epsilon_{\mu\nu\alpha\beta}W^{\alpha}P^{\beta}.
\end{align}
Using these relations it can be shown that $W^{2}$ commutes with $J^{\mu\nu}$ and  $P^{\mu}$. The eigenvalues can easily obtained from
its action on rest frame states. The components of the Pauli-Lubanski four-vector are 
$W_{\mu}=(\bm{J}\cdot \bm{P}, \bm{J}H-\bm{K}\times \bm{P})$ thus acting on rest 
frame states with $W^{2}$ we get the eigenvalues $-p^{2}j(j+1)$. Finally, from the transformation properties of $P^{\mu}$ and $J^{\mu\nu}$ 
under discrete symmetries we can show that $P^{2}$ and $W^{2}$ also commute with discrete symmetries at the quantum level. The 
irreducible representations of the Poincar\'{e} group are labelled by their eigenvalues $p^{2}$ and $-p^{2}j(j+1)$ or, in short, by the mass 
and spin of the free particle.

In the following we will use the simple notation 
\begin{equation}
U({\cal{P}}, 0)\equiv \Pi.
\label{par}
\end{equation}
The symmetries of space-time yield the following CSCO for free elementary particles:  $\{ H,P^{2},W^{2}, J_{3}, \Pi \}$. States 
describing free particles can be labelled by the eigenvalues of this set: $| E, p^{2}, -p^{2}j(j+1), \lambda, \pi \rangle$, which for the 
sake of simplicity we denote as  $| E, p^{2}, j, \lambda, \pi \rangle$ and refer to them as {\it Poincar\'{e} states}. 

 It is very important to remark that the irreducible representations 
of the {\it{proper orthochronous Poincar\'{e} subgroup}} are also are labelled by the eigenvalues of $P^{2}$ and $W^{2}$,  
the good quantum numbers of this subgroup are the eigenvalues of $\{ H,P^{2},W^{2}, J_{3} \}$ and the states in the basis can be 
labelled by their eigenvalues as $| E, p^{2}, -p^{2}j(j+1), \lambda \rangle\equiv| E, p^{2}, j, \lambda \rangle$. The full symmetry group 
of free particles however, is the complete Poincar\'{e} group including discrete symmetries which requires to include parity in the CSCO. 
This means that there must exist states $| E, p^{2}, j, \lambda \rangle$ of both parities $| E, p^{2}, j, \lambda,\pi \rangle$ with $\pi=\pm1$ 
which are the states describing free particles.  
  
It is also worth to remark that $P^{i}$ does not commute with $J_{3}$ nor does commute with parity thus $P^{i}$ cannot be in the CSCO and 
the states  $| E, p^{2}, j, \lambda, \pi \rangle$ are not eigenstates of $\bm{P}$. The Poincar\'{e} good quantum numbers fix only 
$| \bm{p}|^{2}=E^{2}-p^{2}$. Nevertheless,  we 
can always construct free particle states of well defined momentum $\bm{p}$, starting with the states $| E, p^{2}, j, \lambda, \pi \rangle$ 
in the rest frame and using the boost operator on these states. The boost generators $\bm{K}$ commute with $P^{2}$ and $W^{2}$ 
thus a boost will not change the corresponding quantum numbers $p^{2}$ and $j$. Furthermore, the properties of the boost generator 
encoded in the commutators
\begin{equation}
[K^{i}, P^{0}]=i P^{i}, \qquad [K^{i}, P^{j}]=i \delta^{ij}P^{0}, \qquad [K^{i}, \Pi]=2 K^{i}\Pi ,
\end{equation}
causes well defined transformations between the components of the eigenvalues of $P^{\mu}$ leaving $P^{2}$ invariant, thus the 
eigenvalue of $P^{0}$ is defined by those of $P^{2}$ and $P^{i}$.

For massive particles the eigenvalues satisfy $p^{2}=m^{2}$. In the rest frame $E=m$ and rest frame Poincar\'{e} states are 
$ | m, m^{2}, j, \lambda, \pi \rangle $. The rest frame is the only frame where the states 
$| E, p^{2}, j, \lambda, \pi \rangle$ are also eigenstates of $P^{i}$ since $|\bm{p}|^{2}=0$ fixes unambiguously 
$\bm{p}=\bm{0}$, thus if we denote the rest frame momentum as $k^{\mu}=(m,0,0,0)$ we can write 
$ | m, m^{2}, j, \lambda, \pi \rangle  \equiv| k,m^{2}, j, \lambda, \pi\rangle$ where the label $k$ includes the labels of all the components 
of $k^{\mu}$. Rest frame Poincar\'{e} states satisfy
\begin{equation}
P^{\mu}|k,m^{2},j,\lambda,\pi\rangle= k^{\mu} |k,m^{2},j,\lambda,\pi\rangle.
\end{equation}
The classical transformation $L$ taking us from the rest frame four-momentum $k^{\mu}$ to the 
four-momentum $p^{\mu}=(E, p_{x},p_{y},p_{z})$
\begin{equation}
p^{\mu}=L^{\mu}_{~\nu}(p)k^{\nu},
\end{equation}
is represented in the space of quantum states by a boost transformation $U(L(p),0)$.  Using the composition rule of the group it 
is easy to show that
\begin{equation}
P^{\mu}U(L(p),0)=L^{\mu}_{~\alpha}(p)  U(L(p),0)P^{\alpha},
\end{equation}
thus 
\begin{equation}
P^{\mu}U(L(p),0)|k,m^{2},j,\lambda,\pi\rangle=L^{\mu}_{~\alpha}(p) k^{\alpha} U(L(p),0)|k,m^{2},j,\lambda,\pi\rangle
= p^{\mu} U(L(p),0)|k,m^{2},j,\lambda,\pi\rangle.
\end{equation}
Since $P^{2}$ and $W^{2}$ commute with 
$K^{i}=J^{0i}$ the boosted state will have the same eigenvalues $p^{2}$ and $-p^{2}j(j+1)$ and the eigenvalue of $P^{0}$ is now 
$E=\sqrt{|\bm{p}|^{2}+m^{2}}$. However, $K^{i}$ does not commute with $J_{3},\Pi$, thus the boosted state is not eigenstate of 
these operators. With these considerations, the boosted state can be written as
\begin{equation}
|p, m^{2}, j\rangle_{\lambda\pi}=U(L(p),0)|k,m^{2},j,\lambda,\pi\rangle,
\end{equation}
where we use the labels $\lambda\pi$ as subindex in the left hand side of this equation to emphasize that 
$|p, m^{2}, j \rangle_{\lambda\pi}$ is not eigenstate of $J_{3}, \Pi$. This is not a Poincar\'{e} state but it is an eigenstate 
of $P^{\mu}$ constructed from rest frame Poincar\'{e} states.

The conventional result that a general  $L^{\uparrow}_{+}$ transformation on the state $U(L(p),0) |k,m^{2},j,\lambda\rangle$ yields 
a linear combination of all states with momentum $\Lambda p$ obtained from the rest frame states with different polarizations 
$\lambda$ where the coefficients are the matrix elements of the rotation matrix between the rest frame states, is still valid for 
the $ |p,m^{2},j\rangle_{\lambda\pi}$ states   
\begin{align}
U(\Lambda,0)|p, m^{2}, j \rangle_{\lambda\pi}&=
U(L(\Lambda p),0) [U^{-1}(L(\Lambda p),0)U(\Lambda,0)U(L(p),0)] |k,m^{2},j,\lambda, \pi\rangle \nonumber \\
&=U(L(\Lambda p),0)  \sum_{\lambda^{\prime},\pi^{\prime}} | k,m^{2},j,\lambda^{\prime}, \pi^{\prime}\rangle
 \langle k,m^{2},j,\lambda^{\prime}, \pi^{\prime} |
 e^{-i\bm{J}\cdot\bm{\theta}} |k,m^{2},j,\lambda, \pi\rangle \nonumber \\
&=\sum_{\lambda^{\prime}} D^{(j)}_{\lambda^{\prime}\lambda}(R) | \Lambda p, m^{2},j\rangle_{\lambda^{\prime} \pi}.
\label{Wigrot}
\end{align}
Here, $D^{(j)}_{\lambda^{\prime}\lambda}(R)$ are the matrix elements of the rotation operator between the Poincar\'{e} rest frame 
states. These matrix elements are orthogonal in the parity quantum number because $\Pi$ and $\bm{J}$ commute.

The Poincar\'{e} rest frame states are eigenvalues of parity in this frame
\begin{equation}
\Pi(k)  |k,m^{2},j,\lambda,\pi\rangle = \pi |k,m^{2},j,\lambda,\pi\rangle,
\end{equation}
where we use the notation $\Pi(k)$ for the parity operator in the rest frame. We can find how this condition looks in the frame where the 
particle has momentum $p^{\mu}=(E, \bm{p})$ just applying the transformation 
$U(L(p))$. This equation can be rewritten as a condition on the eigenstates of $P^{\mu}$ which reads
\begin{equation}
\left[ U(L(p),0) \Pi(k) U^{-1}(L(p),0) -\pi \right] |p,m^{2},j\rangle_{\lambda,\pi} = 0.
\label{pareq}
\end{equation}
This condition involves the covariant properties of the parity operator and in general we need to find the transformation properties 
of the parity operator under $L^{\uparrow}_{+}$. Since $U(L(p),0) $ is the boost operator for rest frame states 
and from Eq.(\ref{PpropJK}) we get $\Pi U(L(p),0) \Pi^{-1}= U^{-1}(L(p),0)$ we can rewrite this condition as
\begin{equation}
\left[ U^{2}(L(p),0) \Pi(k) -\pi \right] |p,m^{2},j\rangle_{\lambda,\pi} = 0.
\label{pareqbis}
\end{equation}

The states $|p,m^{2},j\rangle_{\lambda,\pi}$ are eigenstates of $P^{\mu}$ with eigenvalue $p^{\mu}$. On the other side, from 
the composition rule of the group we find that every Poincar\'{e} transformation can be factorized into the product of 
a space-time translation and a homogeneous Lorentz transformation (see Eq. (\ref{factorization})), thus in general the states 
can be decomposed into a product of states in the irreps of the Abelian invariant subgroup spanned by the transformations of the 
type $U(\mathbbm{1},a)$ and states in the irreps of the homogeneous Lorentz transformation $U(\Lambda,0)$. The irreps of
the space-time translations invariant subgroup are the eigenstates of $P^{\mu}$ thus  
\begin{equation}
|p,m^{2},j\rangle_{\lambda,\pi} = |p\rangle |p,\bm{\alpha}\rangle_{HLG},
\label{factorized}
\end{equation}  
where $|p,\bm{\alpha}\rangle_{HLG}$ denote states in the irreps of the homogeneous Lorentz group 
characterized by the corresponding good quantum numbers $\bm{\alpha}$ which must contain the information of the 
Poincar\'{e} quantum numbers (beyond $E$). As will be briefly reviewed below, the HLG irreps in general contain several 
Poincar\'{e} irreps and we are faced with the technical problem of isolating the specific states generated by the boost when 
acting on the Poincar\'{e} rest frame states for a given HLG irrep. 

In the following sections we will address this technical problem, show its connection to the 
equations of motion satisfied by spinning particles in the standard model and the relation of parity transformations with gauge 
transformations for fields transforming in the $(1/2,1/2)$ and $(1,0)\oplus (0,1)$ representations in the massless limit.

\section{Irreps of $L^{\uparrow}_{+}$ and parity: kinematical conditions for $(j,0)\oplus (0,j)$}

In order to make this paper as self-contained as possible we briefly review the construction of the irreps of  the proper isochronous 
homogeneous Lorentz subgroup and analyze the action of parity on these subspaces. From  Eq. (\ref{comrelJJ}) we obtain the 
following commutation relations for the generators of this subgroup
\begin{equation}
[J_{i},J_{j}]=i\epsilon_{ijk} J_{k}, \qquad [J_{i},K_{j}]=i\epsilon_{ijk} K_{k}, \qquad [K_{i},K_{j}]=-i\epsilon_{ijk} J_{k}.
\label{HLGalgebra}
\end{equation}
These relations can be rewritten in terms of the operators
\begin{equation}
\bm{A}=\frac{1}{2}(\bm{J} - i \bm{K}), \qquad \bm{B}=\frac{1}{2}(\bm{J} + i \bm{K}),
\end{equation} 
which using (\ref{HLGalgebra}) can be shown to satisfy the $SU(2)_{A}\otimes SU(2)_{B}$ algebra
\begin{equation}
[A_{i},A_{j}]=i\epsilon_{ijk} A_{k}, \qquad [A_{i},B_{j}]=0, \qquad [B_{i},B_{j}]= i\epsilon_{ijk} B_{k}.
\label{ABalgebra}
\end{equation}
The calculation of the irreps of $SU(2)$ are well known. For the $SU(2)_{A}$ they are characterized by the eigenvalues of the 
Casimir operator $\bm{A}^{2}|a,\lambda_{a}\rangle=a(a+1)|a,\lambda_{a}\rangle$ and of the only member of the corresponding 
Cartan subalgebra $A_{3} |a,\lambda_{a}\rangle=\lambda_{a}|a,\lambda_{a}\rangle$ with 
$a=\frac{n}{2}$ and $\lambda_{a}=-a,-a+1,...,a-1,a$; where $n$  is a positive integer or zero. Similar 
relations hold for the irreps of $SU(2)_{B}$ in such a way that the irreps of   $L^{\uparrow}_{+}$ are characterized by two $SU(2)$ 
quantum numbers $(a,b)$ and a basis for these irreps is given by 
\begin{equation}
\{|a,\lambda_{a}\rangle \otimes |b,\lambda_{b}\rangle   \} \equiv \{ |a,b,\lambda_{a},\lambda_{b}\rangle  \}.
\label{basisab}
\end{equation} 
It is clear from 
$\bm{J}=\bm{A}+\bm{B}$ that these irreps contain sectors with $j=|a-b|,|a-b+1|,..., a+b$ as mentioned above; thus the factorization of 
Poincar\'{e} transformations into a product of a space-time translation and an HLG transformation yields reducible 
representations for the proper isochronous Poincar\'{e} group except in the cases $a=0$ or $b=0$ which contain single spin 
$j=b$ and $j=a$ respectively. Since we will be interested in the Poincar\'{e} irreps contained in the irreps of $L^{\uparrow}_{+}$, 
we will in general make a change of basis to the states of well defined $j$ 
\begin{equation}
|a,b; j,\lambda\rangle= M |a,b,\lambda_{a},\lambda_{b}\rangle
\label{Mabjl}
\end{equation}
where $M=\langle a,b\lambda_{a},\lambda_{b}|a,b; j,\lambda \rangle$ is the change of basis matriz whose elements 
are the corresponding Clebsch-Gordon coefficients.

The two Casimir operators for $L^{\uparrow}_{+}$, $\bm{A}^{2}$ and $\bm{B}^{2}$, can be written in terms of  
$J^{\mu\nu}$ and $\tilde{J}^{\mu\nu}=\frac{1}{2}\epsilon^{\mu\nu\alpha\beta}J_{\alpha\beta}$ as follows
\begin{align}
C_{+}&=\bm{A}^{2}+\bm{B}^{2}= \frac{1}{2}(\bm{J}^{2}-\bm{K}^{2})=\frac{1}{4}J^{\mu\nu}J_{\mu\nu}, \label{AplusB}\\
  C_{-}&=\bm{A}^{2}-\bm{B}^{2}=  - i \bm{J}\cdot \bm{K}=\frac{i}{4}\tilde{J}^{\mu\nu}J_{\mu\nu}. 
\label{AminusB}
\end{align} 
These $C_{+}$ and $C_{-}$ combinations are also Casimir operators and we can use thier eigenvalues, $c_{+}=a(a+1)+b(b+1)$ 
and $c_{-}=a(a+1)-b(b+1)$ respectively, to label the irreps of $L^{\uparrow}_{+}$. 

If we now consider the action of parity  on the irreps  of  $L^{\uparrow}_{+}$ according to the composition rule in Eq. (\ref{parpoin}), 
using Eq. (\ref{PpropJK}) we get $\Pi \bm{A}\Pi =\bm{B}$ and $\Pi \bm{B}\Pi =\bm{A}$, thus the $(a,b)$ representation 
space is mapped by parity onto the $(b,a)$ space. The only irreps of $L^{\uparrow}_{+}$ which are also irreps of parity are the 
representations of the type $(a,a)$.
For $b\neq a$ irreps of parity are the $(a,b)\oplus (b,a)$ representations. These irreps of the
homogeneous Lorentz group generate reducible representations for the Poincar\'{e} group when composed according to 
Eq. (\ref{factorization}), except for $a=0$ or $b=0$. We can solve the algebra for the single spin irreps of 
the homogeneous Lorentz group which yield also irreps of the Poincar\'{e} group and allows us to make a matrix representation 
for the operators and the states in Eq. (\ref{pareq}).

\subsection{Solving the $L^{\uparrow}_{+}$ algebra for $(j,0)$ and $(0,j)$.}

Let us consider first the case $a=0$. For these ("left" in the following) representations the states are $SU(2)_{A}$ singlets thus 
$\bm{A}_{L}=\frac{1}{2}(\bm{J}_{L}- i \bm{K}_{L})=\bm{0}$ and $\bm{J}_{L}= i \bm{K}_{L}$.  Using this result we get 
$\bm{B}_{L}=\frac{1}{2}(\bm{J}_{L}+ i \bm{K}_{L})=\bm{J}_{L}$ and the $SU(2)_{B}$ quantum numbers 
coincide with those of the rotation subgroup of the HLG. For this 
reason we denote the left representations as $(0,j)$ and for these representations Eq. (\ref{Mabjl}) reads
\begin{equation}
|0,j; j,\lambda\rangle = |0,j, 0,\lambda\rangle ,
\end{equation} 
thus the matrix $M$ in Eq.(\ref{Mabjl}) is the $(2j+1)\times (2j+1)$ unit matrix.

The representations with  $b=0$ are called "right" representations. In this case we have 
$\bm{B}_{R}=\frac{1}{2}(\bm{J}_{R} + i \bm{K}_{R})=\bm{0}$ i.e. $\bm{J}_{R}= - i \bm{K}_{R}$, which in turn yields 
$\bm{A}_{R}=\frac{1}{2}(\bm{J}_{R}- i \bm{K}_{R})=\bm{J}_{R}$. In this case, it is the $SU(2)_{A}$ subgroup which coincides with 
the rotation subgroup thus we denote these representations as $(j,0)$ . States with well defined angular momentum in 
this case are
\begin{equation}
 |j,0; j,\lambda\rangle = |j,0 ,\lambda,0 \rangle,
\end{equation} 
and  the matrix $M$ in Eq.(\ref{Mabjl}) is also the $(2j+1)\times (2j+1)$ unit matrix. 

States in the $(j,0)$ and $(0,j)$ representations have identical transformation properties under rotations but 
under boosts $(j,0)$ states transforms with the generator $\bm{K}_{R}= i\bm{J}_{R}$ and differ in this respect from the transformation 
properties of $(0,j)$ states which transform with the generator $\bm{K}_{L}= - i\bm{J}_{L}$. 
This difference can be related to the Casimir operators. Indeed, the operator $\bm{A}^{2}+\bm{B}^{2}$ has the same eigenvalue, $j(j+1)$ 
for states in the $(j,0)$ and the $(0,j)$ representations but the eigenvalues of the combination $\bm{A}^{2}-\bm{B}^{2}$ differ in a sign. 
In this concern, it is convenient to define the chirality operator as
\begin{equation}
\chi=\frac{\bm{A}^{2}-\bm{B}^{2}}{a(a+1)+b(b+1)}=\frac{i}{4[a(a+1)+b(b+1)]} \tilde{J}^{\mu\nu}J_{\mu\nu}.
\end{equation}
Under parity we get $\Pi \chi \Pi^{-1}= -\chi$ thus 
\begin{equation}
\{ \Pi, \chi \}=0.
\end{equation}
The eigenvalues $\xi$  of the chirality operator are good quantum numbers of $L^{\uparrow}_{+}$ and we can label the irreps with 
$(a(a+1)+b(b+1),\xi)$ instead of $(a,b)$ in such a way that for the states in the right representations $(j,0)$ corresponds 
to the labels $(j,+1)$, those in the $(0,j)$ representations have the labels $(j,-1)$ and the irreps of the type $(a,a)$ are labelled 
as $(2a(a+1),0)$. Concerning the $(j,0)$ and $(0,j)$ irreps, the label $j$ is redundant and we can denote the states simply as 
$|j,\xi;j,\lambda\rangle\equiv |\xi;j,\lambda\rangle$ such that, in this notation
\begin{align}
|j,0;j,\lambda\rangle &=|+,j,\lambda\rangle, \\
|0,j;j,\lambda\rangle &=|-,j,\lambda\rangle,
\end{align} 
where the label $\pm$ corresponds to the eigenvalue $\xi=\pm 1$. These states have well defined $j$ and $\lambda$ 
but the construction of the states in the irreps of the Poincar\'{e} group must 
have also well defined parity. Parity maps $(j,0)$ onto $(0,j)$ and its action at the level of the basis is
\begin{align}
\Pi(k) |+,j,\lambda\rangle &= |-,j,\lambda\rangle,  \\
\Pi(k) |-,j,\lambda\rangle &= |+,j,\lambda\rangle,
\end{align}
thus the rest frame eigenstates of parity live in the $(j,0)\oplus (0,j)$ representation space and, in the  Poincar\'{e} basis, are given by 
\begin{align}
|k,m^{2},j,\lambda,+\rangle &= \frac{1}{\sqrt{2} }( |+,j,\lambda\rangle + |-,j,\lambda\rangle ), \label{Pstatesp}\\
|k,m^{2},j,\lambda,-\rangle &= \frac{1}{\sqrt{2} }( |+,j,\lambda\rangle - |-,j,\lambda\rangle ). \label{Pstatesm}
\end{align}

The representation of $\bm{J}_{R}$ in the basis $\{ |+;j,\lambda\rangle \}$ for the $(j,0)$ space yields the conventional 
$(2j+1)\times (2j+1)$ matrix representation for the generators of the $j$ irrep of $SU(2)$, $\bm{J}_{R}= \bm{J}^{(j)}$.
Similarly the representation of $\bm{J}_{L}$ in the basis $\{ |-;j,\lambda\rangle \}$ for the $(j,0)$ space yields $\bm{J}_{L}= \bm{J}^{(j)}$.
This allows us to construct explicitly matrix representations for the states in momentum space without any reference to equations 
of motion or Lagrangians. Indeed, for the $(j,0)$ representation space we can start with the representation of the elements in the basis
$\{|+,j,\lambda\rangle\}$ in the rest frame which have the canonical column vector representation of dimension $2j+1$. The corresponding 
vectors in a frame where the particle has momentum $\bm{p}$ is obtained acting with the boost operator on the rest-frame column vectors. 
For this irrep $\bm{K}_{R}= i \bm{J}_{R}$ and the boost operator is given by
\begin{equation}
U_{R}(L(p),0)=e^{-i\bm{K}_{R}\cdot\bm{\phi}}=e^{\bm{J}^{(j)}\cdot\bm{n}\phi}.
\end{equation}
Similarly, for the $(0,j)$ irrep the states in the basis $\{|-,j,\lambda\rangle \}$ are represented by the canonical vectors of dimension $2j+1$ 
in the rest frame. In this case however $\bm{K}_{L}= -i \bm{J}_{L}$ and the boost operator is now given by
\begin{equation}
U_{L}(L(p),0)=e^{-i\bm{K}_{L}\cdot\bm{\phi}}=e^{-\bm{J}^{(j)}\cdot\bm{n}\phi}.
\end{equation}

The calculation of the explicit form of the representation of the states in the frame where the particle has momentum $\bm{p}$ requires 
to explicitly calculate the exponentials and to evaluate the rapidity $\phi$ given by $\cosh\phi=\gamma=E/m$ or 
$\sinh\phi =\gamma \beta= |\bm{p}|/m$. 
The result for the exponentials are different for different values of $j$. It can be obtained using that the eigenvalues of 
$h\equiv \bm{J}^{(j)}\cdot \bm{n}$ are the same as those of $J^{(j)}_{z}$ i.e. $\lambda=-j, -j+1,...,j-1,j$. On the other side, every 
matrix satisfies the characteristic equation of their eigenvalues i.e. 
\begin{equation}
(h+j)(h+j-1)(h+j-2)...(h-j-1)(h-j)=0.
\end{equation}
This condition reduces the powers of $h$ in the expansion of the exponential and in general we obtain a polynomial of $h$ with 
$\phi$-dependent coefficients. In general for arbitrary $j$ we will obtain a polynomial in $h$ of order $2j$ and the calculation of the 
boost operators for the $(j,0)$ and $(0,j)$ representations is straightforward. The representation of the states in a frame where the 
particle has momentum $p^{\mu}$ can be obtained boosting the representation of the rest frame  states. 

The construction of the states in the irreps of  the Poincar\'{e} group requires to consider the $(j,0)\oplus (0,j)$ space. A basis for this
space is given by  $\{ |+,j,\lambda \rangle, |-,j,\lambda \rangle \}$. The representation of the states in this basis is also canonical but 
now in a space of dimension $2(2j+1)$. In this basis, the chirality operator  and the parity operator have the following matrix 
representations
\begin{align}
\chi= \begin{pmatrix}  \mathbbm{1}_{2j+1} & 0\\ 0&- \mathbbm{1}_{2j+1}  \end{pmatrix}, \qquad 
\Pi (k)=\begin{pmatrix} 0 & \mathbbm{1}_{2j+1} \\ \mathbbm{1}_{2j+1} & 0 \end{pmatrix}.
\end{align}
The representation of the Poincar\'{e} rest frame states in Eq. (\ref{Pstatesp},\ref{Pstatesm}) can be obtained from the representation 
of the states  $ |+,j,\lambda \rangle$ and  $|-,j,\lambda \rangle \}$. 
The proper orthochronous HLG transformations in this space are block diagonal
\begin{equation}
U({\Lambda},0)=\begin{pmatrix}U_{R}(\Lambda,0)&0\\0&U_{L}(\Lambda,0) \end{pmatrix},
\end{equation}
and the matrix representation for the generators read
\begin{align}
\bm{J}=\begin{pmatrix} \bm{J}_{R}&0\\0&\bm{J}_{L} \end{pmatrix}=\begin{pmatrix} \bm{J}^{(j)}&0\\0&\bm{J}^{(j)} \end{pmatrix}, \qquad 
\bm{K}=\begin{pmatrix} \bm{K}_{R}&0\\0&\bm{K}_{L} \end{pmatrix}=i \begin{pmatrix}  \bm{J}^{(j)}&0\\0&-\bm{J}^{(j)} \end{pmatrix}. 
\end{align}

In order to illustrate the process in the remaining of this section we explicitly work out the cases $j=1/2$ and $j=1$.

\subsection{The case $j=\frac{1}{2}$}

In this case the matrix representation for the angular momentum generators of the $(1/2,0)$ representation in the basis 
$\{|+,1/2,\lambda \rangle\}$ are $\bm{J}_{R}=\bm{J}^{(\frac{1}{2})}=\bm{\sigma}/2 $ where $\bm{\sigma}$ are the Pauli matrices. Similarly, 
the matrix representation for the angular momentum generators of the $(0,1/2)$ representation in the basis 
$\{|-,1/2,\lambda \rangle\}$ are $\bm{J}_{L}=\bm{J}^{(\frac{1}{2})}=\bm{\sigma}/2 $. The operator $h=\bm{J}^{(\frac{1}{2})}\cdot\bm{n}$ 
satisfies
\begin{equation}
(h+\frac{1}{2})(h-\frac{1}{2})=0 \Rightarrow h^{2} = \frac{1}{4}
\end{equation}
and we obtain
\begin{align}
U_{R}(L(p),0)&=e^{h\phi}= \cosh\frac{\phi}{2} + 2h \sinh\frac{\phi}{2} =\frac{E+m+\bm{\sigma}\cdot\bm{p}}{\sqrt{2m(E+m)}}, \\
U_{L}(L(p),0)&=e^{-h\phi}= \cosh\frac{\phi}{2} - 2h \sinh\frac{\phi}{2} =\frac{E+m-\bm{\sigma}\cdot\bm{p}}{\sqrt{2m(E+m)}}.
\end{align} 

The representation of the rest frame states in $(\frac{1}{2},0)$ in the basis $\{ |+,1/2, \lambda \rangle  \}$ is canonical and the two 
component right spinors are obtained acting with $U_{R}(L(p))$ on them. Similarly for the states in the left representation 
$(0,\frac{1}{2})$. These spinors however cannot describe elementary free particles because they are not parity eigenstates. 
The construction of states with well defined parity requires to work with the $(\frac{1}{2},0)\oplus (0,\frac{1}{2})$ space. The matrix 
representation of the generators, chirality  and parity operators are  
\begin{equation}
\bm{J}=\frac{1}{2}\begin{pmatrix}\bm{\sigma}&0\\0&\bm{\sigma} \end{pmatrix}, \qquad
\bm{K}=\frac{i}{2}\begin{pmatrix}\bm{\sigma}&0\\0& -\bm{\sigma} \end{pmatrix}, \qquad 
\chi=\begin{pmatrix}\mathbbm{1}&0\\0& -\mathbbm{1}\end{pmatrix}, \qquad
\Pi (k)=\begin{pmatrix} 0 & \mathbbm{1} \\ \mathbbm{1} & 0 \end{pmatrix}.
\end{equation}
Notice that $\chi$ and $\Pi$ do not commute, thus free particle states cannot have well defined chirality.
In the rest frame, the representation of the Poincar\'{e} states $|k,m^{2},j,\lambda,\pi \rangle$ are denoted as 
$\phi(k,m^{2},j,\lambda,\pi)$. These four-component spinors satisfy   
\begin{align}
\Pi (k) \phi(k,m^{2},j,\lambda,\pi) &= \pi \phi(k,m^{2},j,\lambda,\pi)  \label{pareq12}\\
J_{3} \phi(k,m^{2},j,\lambda,\pi) &= \lambda \phi(k,m^{2},j,\lambda,\pi).
\end{align} 
The simultaneous solution to these equations yields the following representation of the states in Eqs. (\ref{Pstatesp},\ref{Pstatesm})
\begin{align}
\phi(k,m^{2},\frac{1}{2},\frac{1}{2},+)&= \frac{1}{\sqrt{2}} \begin{pmatrix}1\\0\\1\\0 \end{pmatrix}, \quad
\phi(k,m^{2},\frac{1}{2},-\frac{1}{2},+)= \frac{1}{\sqrt{2}} \begin{pmatrix}0\\1\\0\\1 \end{pmatrix}, \nonumber \\
\phi(k,m^{2},\frac{1}{2},\frac{1}{2},-)&=\frac{1}{\sqrt{2}}  \begin{pmatrix}1\\0\\-1\\0 \end{pmatrix}, \quad
\phi(k,m^{2},\frac{1}{2},-\frac{1}{2},-)= \frac{1}{\sqrt{2}} \begin{pmatrix}0\\1\\0\\-1 \end{pmatrix}.
\end{align} 

The boost operator for this representation is given by
\begin{align}
U(L(p),0) &
=\frac{1}{\sqrt{2m(E+m)}}
\begin{pmatrix} 
E+m+p_{z} & p_{x} - ip_{y} & 0 & 0\\ 
p_{x} + ip_{y} & E+m-p_{z} & 0 & 0 \\
0&0& E+m- p_{z} &- p_{x} + ip_{y} \\
0&0& - p_{x} - ip_{y} & E+m+p_{z} 
\end{pmatrix} .
\label{Diracboost}
\end{align}

The states $|p,m^{2},j\rangle_{\lambda\pi}$ are represented by four-component spinors denoted as 
$\phi_{\lambda\pi}(p,m^{2},j)$ and are obtained using the boost operator on the corresponding 
rest frame spinors. We obtain
\begin{align}
\phi_{\frac{1}{2}+}(p,m^{2},\frac{1}{2})&= N \begin{pmatrix}E+m+p_{z}\\p_{x} + ip_{y}\\E+m- p_{z}\\ - p_{x} - ip_{y} \end{pmatrix}, \qquad
\phi_{-\frac{1}{2}+}(p,m^{2},\frac{1}{2})=N  \begin{pmatrix} p_{x} - ip_{y}\\ E+m-p_{z} \\ - p_{x} + ip_{y} \\ E+m+p_{z} \end{pmatrix}, \\
\phi_{\frac{1}{2}-}(p,m^{2},\frac{1}{2})&=N  \begin{pmatrix}E+m+p_{z}\\p_{x} + ip_{y}\\ -E- m + p_{z}\\  p_{x} + ip_{y} \end{pmatrix}, \qquad
\phi_{-\frac{1}{2}-}(p,m^{2},\frac{1}{2})= N \begin{pmatrix} p_{x} - ip_{y}\\ E+m-p_{z} \\  p_{x} - ip_{y} \\ -E-m-p_{z} \end{pmatrix},
\end{align} 
where $N=1/\sqrt{2m(E+m)}$. Notice that these spinors satisfy non-trivial conditions related to the fact that in the rest frame they 
have well defined parity and $J_{3}$. In particular, the parity condition in the frame where the particles has momentum $p^{\mu}$ 
is obtained just boosting the eigenvalue Eq.(\ref{pareq12}) to obtain 
\begin{align}
U(L(p),0)\Pi (k) U^{-1}(L(p),0)\phi_{\lambda\pi}(p,m^{2},\frac{1}{2}) &= \pi \phi_{\lambda\pi}(p,m^{2},\frac{1}{2})  .
\end{align} 
A straightforward calculation using Eq.(\ref{Diracboost}) yields
\begin{align}
U(L(p),0)\Pi(k) U^{-1}(L(p),0) = \frac{\gamma^{\mu}p_{\mu}}{m} ,
\end{align} 
where
\begin{align}
\gamma^{0}=\Pi(k), \qquad \gamma^{i} = \begin{pmatrix}0&-\sigma^{i}\\ \sigma^{i} &0 \end{pmatrix}.
\end{align}
Using this result, the parity eigenvalue equation in the frame where te particle has momentum $p^{\mu}$ reads
\begin{align}
(\gamma^{\mu}p_{\mu} -\pi m) \phi_{\lambda\pi}(p,m^{2},\frac{1}{2}) =0.
\end{align} 
These are the conditions for particle ($\pi=+1$) and anti-particle ($\pi=-1$) solutions of the Dirac equation and our first principles 
calculation reveals that Dirac equation is just the covariant form of the parity eigenvalue equation. Notice that $\gamma^{\mu}$ appear 
here as the covariant companions of the parity operator. Also, if we define
\begin{equation}
\Pi (p)=U(L(p),0)\Pi (k) U^{-1}(L(p),0),
\end{equation}
it is clear that this operator satisfies $\Pi^{2} (p)=\mathbbm{1}$, thus the relation
\begin{equation}
(\gamma^{\mu}p_{\mu})^{2}=p^{2},
\end{equation}
must hold.

\subsection{The case $j=1$}

In the case $j=1$ the representation of $\bm{J}_{R}$ in the basis $\{|+,1,\lambda\rangle\}$ is $\bm{J}_{R}=\bm{J}^{(1)}$. Similarly,
 the representation of $\bm{J}_{L}$ in the basis $\{|-,1,\lambda\rangle\}$ is $\bm{J}_{L}=\bm{J}^{(1)}$ where
\begin{align}
J^{(1)}_{1}=\begin{pmatrix} 0&\frac{1}{\sqrt{2}}&0 \\ \frac{1}{\sqrt{2}}&0&\frac{1}{\sqrt{2}}\\0&\frac{1}{\sqrt{2}}&0 \end{pmatrix}, \qquad
J^{(1)}_{2}=\begin{pmatrix} 0&-\frac{i}{\sqrt{2}}&0 \\ \frac{i}{\sqrt{2}}&0&-\frac{i}{\sqrt{2}}\\0&\frac{i}{\sqrt{2}}&0 \end{pmatrix}, \qquad
J^{(1)}_{3}=\begin{pmatrix} 1&0&0 \\0&0&0\\0&0&-1 \end{pmatrix}.
\label{J1}
\end{align}
In this case $h$ satisfies
\begin{equation}
(h+1)(h-0)(h+1)=0 \Rightarrow h^{3} = h
\end{equation}
and for the boost operators of the  $(1,0)$ and $(0,1)$ representations we obtain
\begin{align}
U_{R}(L(p),0)&=e^{h\phi}= 1+ h \sinh\phi  + h^{2} (\cosh\phi -1 )
= 1 + \frac{\bm{J}^{(1)}\cdot\bm{p}}{m} + \frac{ (\bm{J}^{(1)}\cdot\bm{p})^{2}}{m(E+m)} , \\
U_{L}(L(p),0)&=e^{-h\phi}=1 - h \sinh\phi  + h^{2} (\cosh\phi -1 )
=1 - \frac{\bm{J}^{(1)}\cdot\bm{p}}{m} + \frac{ (\bm{J}^{(1)}\cdot\bm{p})^{2}}{m(E+m)},
\end{align} 
such that the boost operator of the $(1,0)\oplus (0,1)$ space is
\begin{align}
U(L(p),0)&
= N \begin{pmatrix} m(E+m) + (E+m) \bm{J}^{(1)}\cdot\bm{p} + (\bm{J}^{(1)}\cdot\bm{p})^{2}& 0 \\
0 &  m(E+m) - (E+m) \bm{J}^{(1)}\cdot\bm{p} + (\bm{J}^{(1)}\cdot\bm{p})^{2} \end{pmatrix} ,
\label{B1001}
\end{align}
where $N= 1/m(E+m)$.

States with well defined parity and $J_{3}$ in the rest frame are described by the following six-component 
$\phi(k,m^{2},j,\lambda,\pi)$ column vectors
\begin{align}
\phi(k,m^{2},1,1,+)&= \begin{pmatrix}1\\0\\0\\1\\0\\0 \end{pmatrix}, \qquad
\phi(k,m^{2},1,0,+)= \begin{pmatrix}0\\1\\0\\0\\1\\0 \end{pmatrix}, \qquad
\phi(k,m^{2},1,-1,+)= \begin{pmatrix}0\\0\\1\\0\\0\\1 \end{pmatrix},  \\
\phi(k,m^{2},1,1,-)&= \begin{pmatrix}1\\0\\0\\-1\\0\\0 \end{pmatrix}, \qquad
\phi(k,m^{2},1,0,-)= \begin{pmatrix}0\\1\\0\\0\\-1\\0 \end{pmatrix}, \qquad
\phi(k,m^{2},1,-1,-)= \begin{pmatrix}0\\0\\1\\0\\0\\-1 \end{pmatrix}.
\end{align} 
These spinors satisfy the parity eigenvalue equation
\begin{equation}
\Pi (k)\phi(k,m^{2},1,\lambda,\pi)= \pi \phi(k,m^{2},1,\lambda,\pi).
\label{pee1}
\end{equation}
In the frame where the particle has momentum $p^{\mu}$ are given by the following $\phi_{\lambda\pi}(p,m^{2},j)$ spinors
\begin{align}
\phi_{1+}(p,m^{2},1)&= N \begin{pmatrix}(E+m)(m+p_{z})+ p^{2}_{z} +p_{+}p_{-} \\ (E+m+p_{z})p_{+}\\p^{2}_{+}\\
(E+m)(m-p_{z})+ p^{2}_{z} +p_{+}p_{-} \\-(E+m - p_{z})p_{+}\\p^{2}_{+} \end{pmatrix}, \qquad
\phi_{0+}(p,m^{2},1)= N \begin{pmatrix}(E+m+p_{z})p_{-}\\ m(E+m)+2p_{+}p_{-}\\(E+m-p_{z})p_{+}\\-(E+m-p_{z})p_{-} 
\\m(E+m)+2p_{+}p_{-}\\ -(E+m+p_{z})p_{+} \end{pmatrix}, \nonumber \\
\phi_{-1+}(p,m^{2},1)&= N \begin{pmatrix} p^{2}_{-}\\(E+m-p_{z})p_{-}\\ (E+m)(m-p_{z})+p^{2}_{z}+p_{+}p_{-} \\
p^{2}_{-}\\-(E+m+p_{z})p_{-}\\ (E+m)(m+p_{z})+p^{2}_{z}+p_{+}p_{-} \end{pmatrix},  \quad
\phi_{1-}(p,m^{2},1)= N \begin{pmatrix}(E+m)(m+p_{z})+ p^{2}_{z} +p_{+}p_{-} \\ (E+m+p_{z})p_{+}\\p^{2}_{+}\\
-(E+m)(m-p_{z})- p^{2}_{z} -p_{+}p_{-} \\(E+m - p_{z})p_{+}\\ - p^{2}_{+} \end{pmatrix}, \nonumber \\
\phi_{0-}(p,m^{2},1)&= N \begin{pmatrix}(E+m+p_{z})p_{-}\\ m(E+m)+2p_{+}p_{-}\\(E+m-p_{z})p_{+}\\(E+m-p_{z})p_{-} 
\\ - m(E+m)- 2p_{+}p_{-}\\ (E+m+p_{z})p_{+} \end{pmatrix}, \qquad
\phi_{-1-}(p,m^{2},1)= N  \begin{pmatrix} p^{2}_{-}\\(E+m-p_{z})p_{-}\\ (E+m)(m-p_{z})+p^{2}_{z}+p_{+}p_{-} \\
- p^{2}_{-}\\(E+m+p_{z})p_{-}\\ -(E+m)(m+p_{z})-p^{2}_{z}-p_{+}p_{-} \end{pmatrix}.
\end{align} 
 
 The covariant form of parity operator is obtained using Eq.(\ref{B1001})
 \begin{align}
U(L(p),0)\Pi (k) U^{-1}(L(p),0) = \frac{S^{\mu\nu}p_{\mu}p_{\nu}}{m^{2}} ,
\end{align} 
where $S^{\mu\nu}$ are $6\times 6$ symmetric traceless matrices satisfying $S^{\mu}_{~\mu}=0$ which leaves only nine independent 
matrices. These matrices can be written in terms of parity and the generators of the HLG as \cite{Gomez-Avila:2013qaa}
\begin{equation}
S_{\mu\nu}=\Pi(k) \left( g_{\mu\nu} -i (g_{0\mu}M_{0\nu}+g_{0\nu}M_{0\mu}) -\{ M_{0\mu}, M_{0\nu} \} \right)
\end{equation}
The explicit form of these matrices in terms of parity, chirality and the generators of rotations is
\begin{align}
S^{00}=\Pi(k), \qquad S^{0i}=-\Pi(k) \chi J^{i} ,\qquad S^{ij}= \Pi(k) (-\delta^{ij} +\{J^{i},J^{j} \} ) .
\end{align}
Boosting Eq. (\ref{pee1}) we obtain the covariant form of the parity eigenvalue condition as
\begin{align}
[S^{\mu\nu}p_{\mu}p_{\nu}- \pi m^{2} ] \phi_{\lambda\pi}(p,m^{2},1)=0.
\label{peec1}
\end{align} 
This condition was obtained by Weinberg long ago \cite{Weinberg:1964cn}, and the corresponding equation in configuration space
\begin{align}
[S^{\mu\nu}\partial_{\mu}\partial_{\nu} +  m^{2} ] \psi (x)=0.
\label{peec1cs}
\end{align} 
suggested as the equation of motion for the corresponding field. However, later on it was shown that this equation of motion, in 
addition to the solutions $\psi(x)=e^{-i p\cdot x}\phi_{\lambda\pi}(p,m^{2},1)$ has also solutions with $p^{2}=-m^{2}$ 
(tachyonic solutions). This can be traced to the algebraic properties of $S_{\mu\nu}$ studied in Ref. \cite{Gomez-Avila:2013qaa} 
which yield
\begin{equation}
(S^{\mu\nu}p_{\mu}p_{\nu})^{2} \equiv (S(p))^{2}=p^{4} \mathbbm{1},
\end{equation} 
such that multiplying on the left Eq.(\ref{peec1cs}) with $S(\partial)-m^{2}$ we get
\begin{align}
[\partial^{4}- m^{4} ]\psi(x)=0.
\end{align} 
The problem can be solved if we take into account that parity and $p^{2}$ are independent quantum numbers. Said in other words, 
the parity eigenvalue equation must be imposed independently of the $p^{2}=m^{2}$ condition. The simplest way to do this is to use 
projectors onto well defined parity and $p^{2}$. 

The projectors onto well defined parity ($\pi=\pm$) subspaces, in the rest frame, read%
\begin{equation}
\mathbb{P}_{\pi}(k)=\frac{1}{2}(\mathbbm{1}+\pi\Pi (k)). 
\end{equation}
In the frame where he particle has momentum $p$ these projectors take the form
\begin{equation}
\mathbb{P}_{\pi}(p)=\frac{1}{2} (\mathbbm{1} + \pi \frac{S(p)}{p^{2}}).
\end{equation} 
It is easy to show that these operators satisfy the projection conditions
\begin{equation}
\mathbb{P}^{2}_{\pm}(p)=\mathbb{P}_{\pm}(p), \qquad \mathbb{P}_{+}(p)\mathbb{P}_{-}(p)=\mathbb{P}_{-}(p)\mathbb{P}_{+}(p)=0,
\qquad \mathbb{P}_{+}(p)+\mathbb{P}_{-}(p)=\mathbbm{1}.
\end{equation} 
The simultaneous projection onto well defined (positive) parity and Poincar\'e orbit yields the condition
\begin{equation}
\frac{p^{2}}{m^{2}}\mathbb{P}_{-}(p)\phi_{\lambda\pi}(p,m^{2},1)=\phi_{\lambda\pi}(p,m^{2},1)
\end{equation} 
which can be rewritten as
\begin{equation}
(\Sigma_{\mu\nu}p^{\mu}p^{\nu} - m^{2} \mathbbm{1} )\phi_{\lambda\pi}(p,m^{2},1)=0.
\label{eomj1}
\end{equation}  
where  $\Sigma_{\mu\nu}=\frac{1}{2}(g_{\mu\nu}+S_{\mu\nu})$. 
In configuration space the spinor $\psi(x)=e^{-i p\cdot x}\phi_{\lambda\pi}(p,m^{2},1)$ satisfy the equation
\begin{equation}
(\Sigma_{\mu\nu}\partial^{\mu}\partial^{\nu} + m^{2} \mathbbm{1} )\psi(x)=0.
\end{equation}  
This equation can be derived from the Hermitian Lagrangian
\begin{equation}
{\cal L}_{0}=\partial^{\mu} \bar{\psi}\Sigma_{\mu\nu}\partial^{\nu}\psi - m^{2} \bar{\psi}\psi.
\label{L0}
\end{equation}  
Further details on the formal structure of the field theory and its canonical quantization can be found in 
\cite{Gomez-Avila:2013qaa,Napsuciale:2015kua}. 
The possibility that dark matter fields transforms in this representation of the HLG and the phenomenology of the leading 
terms in the corresponding effective theory for their interactions with standard model fields in a hidden scenario can be 
found in \cite{Hernandez-Arellano:2018sen, Hernandez-Arellano:2019qgd}. Here we are concerned with the ultraviolet 
behavior on the view of the canonical dimension one of this field which can be read from Eq.(\ref{L0}) and in turn yields dimension 
four for the leading terms of the effective theory \cite{Hernandez-Arellano:2018sen}. This observation opens the possibility 
that this fields be useful in the model building of extensions of the standard model with dark matter in a hidden scenario where 
the bridge between the standard model and the dark world be given by the dimension four terms constructed from singlets on 
both sides. There are only three dimension four terms, one of them violating parity. If we put aside this last term, after the 
phenomenology worked out in \cite{Hernandez-Arellano:2018sen, Hernandez-Arellano:2019qgd} it is clear that all studied 
observables are consistent with only one of the operators, $\bar{\psi}\psi\phi^{\dagger}\phi$, which yields a Higgs portal to 
dark matter in this representation. Since fields transforming in the $(1,0)\oplus (0,1)$ representation are conventionally described 
with an antisymmetric rank two tensor (see section V for this description) we call tensor dark matter (TDM) to this field. In order 
to clarify if there is a chance of having a sensible fundamental theory we must study the possibilities for TDM interactions and 
mass generation mechanisms which may yield a good ultraviolet behavior. 
The possible bad ultraviolet behavior can be identified in the propagator which for TDM was obtained in \cite{Napsuciale:2015kua} and reads
\begin{equation}
iS_{T}(p)=\frac{-p^{2}+S(p)+2m^{2}}{2m^{2}(p^{2}-m^{2}+i\epsilon)}.
\end{equation}
Clearly, this propagator yield amplitudes which violate unitarity in the ultraviolet for the interacting theory. 
In order to exhibit the origin of this bad 
behavior at high energies we rewrite the propagator in terms of the parity projectors to obtain
\begin{equation}
iS_{T}(p)=\frac{\mathbb{P}_{+}(p)-\frac{p^{2}-m^{2}}{m^{2}}\mathbb{P}_{-}(p)}{p^{2}-m^{2}+i\epsilon}.
\label{proptensor}
\end{equation}
The problematic terms are clearly related to unphysical negative parity terms which vanish on-shell but get active 
off-shell in the interacting theory. This situation is very similar to the case of the massive vector particle, i.e. for the spin-one 
particles transforming in the $(\frac{1}{2},\frac{1}{2})$ representation as we will show in the next section.

\section{Solving the algebra for the {\it{quantum}} $(\frac{1}{2},\frac{1}{2})$ representation.}

The $(\frac{1}{2},\frac{1}{2})$ representation is the lowest irrep of $L^{\uparrow}_{+}$ with more than one Poincar\'{e} sectors and requires
to devise a procedure to eliminate the undesired (unphysical) sector from the theory. It is also  the lowest $L^{\uparrow}_{+}$ 
representation which is also an irrep of parity. Here we will do a first principles construction of this representation 
at the quantum level.

\subsection{ $(\frac{1}{2},\frac{1}{2})$ as tensor product of fundamental representations}
For every Lie group higher dimensional representations can be constructed from the products of the fundamental representations. 
For the $(\frac{1}{2},\frac{1}{2})$ representation we have two possibilities: $(\frac{1}{2},0)\otimes (0,\frac{1}{2})$ and 
$(0,\frac{1}{2})\otimes (\frac{1}{2},0)$.

A basis for $V_{RL}\equiv (\frac{1}{2},0)\otimes (0,\frac{1}{2})$ is induced by the basis of  
$(\frac{1}{2},0)$and $(0,\frac{1}{2})$. It consist of the following tensor products 
$\{|+,\frac{1}{2},\lambda\rangle \otimes|-,\frac{1}{2}\lambda^{\prime}\rangle \}$. Similarly, a basis for 
$V_{LR}\equiv (0,\frac{1}{2})\otimes (\frac{1}{2},0)$ is given by $\{|-,\frac{1}{2},\lambda\rangle \otimes|+,\frac{1}{2}\lambda^{\prime}\rangle \}$. 
We will call in the following these bases as tensor product (TP) bases.
A general $L^{\uparrow}_{+}$ transformation in $V_{RL}$ can be written as 
\begin{equation}
U^{TP}_{RL}(\Lambda,0)=U^{TP}_{R}(\Lambda,0) \otimes U^{TP}_{L}(\Lambda,0). 
\label{BRL}
\end{equation}
The corresponding generators read%
\begin{equation}
\bm{J}^{TP}_{RL}=\frac{1}{2}(\bm{\sigma}\otimes \mathbbm{1}+\mathbbm{1}\otimes\bm{\sigma}), \qquad 
\bm{K}^{TP}_{RL}=\frac{i}{2}(\bm{\sigma}\otimes\mathbbm{1}-\mathbbm{1}\otimes\bm{\sigma}).
\end{equation}
These operators satisfy
\begin{equation}
(\bm{J}^{TP}_{RL}\cdot\bm{n})^{3}=\bm{J}^{TP}_{RL}\cdot\bm{n}, \qquad (i\bm{K}^{TP}_{RL}\cdot\bm{n})^{3}=i\bm{K}^{TP}_{RL}\cdot\bm{n},
\label{JKTPV}
\end{equation}
which allows us to solve the algebra and to obtain explicit expressions for the rotations and boosts operators. In particular the 
boost operator for rest frame states can be obtained either exponentiating the generators and using 
the relations in Eq. (\ref{JKTPV}) which yields
\begin{equation}
U^{TP}_{RL}(L(p),0)=\mathbbm{1} - \sinh \phi (i\bm{K}^{TP}_{RL} \cdot \bm{n})+ (\cosh\phi-1) (i\bm{K}^{TP}_{RL}\cdot \bm{n})^{2},
\end{equation}
or directly as the tensor product of the boost operator for the $(\frac{1}{2},0)$ and $(0,\frac{1}{2})$ representations
\begin{equation}
U^{TP}_{RL}(L(p),0)=\frac{1}{2m(E+m)}[E+m+\bm{\sigma}\cdot\bm
{p}]\otimes\lbrack E+m-\bm{\sigma}\cdot\bm{p}].
\end{equation}
We obtain
\begin{equation}
U^{TP}_{RL}(L(p),0)= \frac{1}{2m(E+m)}
\left(\begin{array}[c]{cccc}%
(E+m)^{2}-p_{z}^{2} & -(E+m+p_{z})p_{-} & (E+m-p_{z})p_{-} & -p^{2}_{-}\\
-(E+m+p_{z})p_{+} & (E+m+p_{z})^{2} & -p_{+}p_{-} & (E+m+p_{z})p_{-}\\
(E+m-p_{z})p_{+} & -p_{+}p_{-}  & (E+m-p_{z})^{2} & - (E+m-p_{z})p_{-}\\
-p^{2}_{+} & (E+m+p_{z})p_{+} &- (E+m-p_{z})p_{+}  & (E+m)^{2}-p_{z}^{2} 
\end{array}
\right) ,
\end{equation}
where $p_{\pm}=p_{x}\pm i p_{y}$.

As for the $V_{LR}$ space, a general $L^{\uparrow}_{+}$ transformation can be written as 
\begin{equation}
U^{TP}_{LR}(\Lambda,0)=U^{TP}_{L}(\Lambda,0) \otimes U^{TP}_{R}(\Lambda,0). 
\label{BLR}
\end{equation}
The corresponding generators read%
\begin{equation}
\bm{J}^{TP}_{LR}=\frac{1}{2}(\bm{\sigma}\otimes \mathbbm{1}+\mathbbm{1}\otimes\bm{\sigma}), \qquad 
\bm{K}^{TP}_{LR}=-\frac{i}{2}(\bm{\sigma}\otimes\mathbbm{1}-\mathbbm{1}\otimes\bm{\sigma}).
\end{equation}
Notice that states in $V_{RL}$ have the same transformation properties under rotations as states in the $V_{LR}$ space but different
transformation properties under boosts
\begin{equation}
\bm{J}^{TP}_{RL}= \bm{J}^{TP}_{LR}, \qquad \bm{K}^{TP}_{RL}= -\bm{K}^{TP}_{LR}.
\end{equation}

Under parity $V_{RL}\leftrightarrow V_{LR}$, thus the implementation of parity as a good quantum number requires to consider 
both $V_{RL}$ and $V_{LR}$ versions of the $(1/2,1/2)$ representation which double the number of degrees of freedom with 
respect to its classical version.  The irreducible space for parity is $V=V_{RL}\oplus V_{LR}$.  A basis for this space is given by 
\begin{equation}
\{ |+,\frac{1}{2},\lambda\rangle \otimes|-,\frac{1}{2}\lambda^{\prime}\rangle, \,
|-,\frac{1}{2},\lambda \rangle \otimes|+,\frac{1}{2}\lambda^{\prime}\rangle \}.
\end{equation}
The action of parity on the states in the basis is given by 
\begin{align}
\Pi (k) \left(|+,\frac{1}{2},\lambda\rangle \otimes|-,\frac{1}{2}\lambda^{\prime}\rangle \right)&= 
|-,\frac{1}{2},\lambda \rangle \otimes|+,\frac{1}{2}\lambda^{\prime}\rangle , \\
\Pi (k) \left(|-,\frac{1}{2},\lambda\rangle \otimes|+,\frac{1}{2}\lambda^{\prime}\rangle \right)&= 
|+,\frac{1}{2},\lambda \rangle \otimes|-,\frac{1}{2}\lambda^{\prime}\rangle 
\label{parvec}
\end{align}
thus the matrix representation of parity in this basis is the $8\times 8$ matrix
\begin{equation}
\Pi^{TP}(k)=\begin{pmatrix} 0 &\mathbbm{1} \\ \mathbbm{1}&0 \end{pmatrix}.
\end{equation}

The representation  of $L^{\uparrow}_{+}$ transformations in $V$ is block diagonal
\begin{equation}
U^{TP}(\Lambda,0)=\begin{pmatrix} U^{TP}_{RL}(\Lambda,0) & 0 \\0 & U^{TP}_{LR}(\Lambda,0) \end{pmatrix},
\end{equation}
and the generators for rotations and boost read
\begin{equation}
\bm{J}^{TP}_{V}=\begin{pmatrix} \bm{J}^{TP}_{RL} & 0 \\ 0 & \bm{J}^{TP}_{LR} \end{pmatrix}, \qquad
\bm{K}^{TP}_{V}=\begin{pmatrix} \bm{K}^{TP}_{RL} & 0 \\ 0 & \bm{K}^{TP}_{LR} \end{pmatrix}
=\begin{pmatrix} \bm{K}^{TP}_{RL} & 0 \\ 0 & -\bm{K}^{TP}_{RL} \end{pmatrix}.
\end{equation}

States in a frame where the particle has momentum $p^{\mu}$ can be obtained boosting the rest frame states.
However, rest frame states in the tensor product basis have no well defined $W^{2}$ (total angular momentum). 
States with well defined total angular momentum in Eq. (\ref{Mabjl}) are obtained in the rest frame with the following 
change of basis 
\begin{equation}
\left(
\begin{array}
[c]{c}%
|\frac{1}{2},\frac{1}{2}; 0,0\rangle_{RL}\\
|\frac{1}{2},\frac{1}{2}; 1,1\rangle_{RL}\\
|\frac{1}{2},\frac{1}{2}; 1,0\rangle_{RL}\\
|\frac{1}{2},\frac{1}{2}; 1,-1\rangle_{RL}\\
|\frac{1}{2},\frac{1}{2}; 0,0\rangle_{LR}\\
|\frac{1}{2},\frac{1}{2}; 1,1\rangle_{LR}\\
|\frac{1}{2},\frac{1}{2}; 1,0\rangle_{LR}\\
|\frac{1}{2},\frac{1}{2}; 1,-1\rangle_{LR}
\end{array}
\right)  = M_{A}
 \left(
\begin{array}
[c]{c}%
|+,\frac{1}{2},\frac{1}{2} \rangle \otimes |-,\frac{1}{2},\frac{1}{2} \rangle \\
|+,\frac{1}{2},\frac{1}{2} \rangle \otimes |-,\frac{1}{2},-\frac{1}{2} \rangle\\
|+,\frac{1}{2},-\frac{1}{2} \rangle \otimes |-,\frac{1}{2},\frac{1}{2} \rangle\\
|+,\frac{1}{2},-\frac{1}{2} \rangle \otimes |-,\frac{1}{2},-\frac{1}{2} \rangle\\
|-,\frac{1}{2},\frac{1}{2} \rangle \otimes |+,\frac{1}{2},\frac{1}{2} \rangle \\
|-,\frac{1}{2},\frac{1}{2} \rangle \otimes |+,\frac{1}{2},-\frac{1}{2} \rangle\\
|-,\frac{1}{2},-\frac{1}{2} \rangle \otimes |+,\frac{1}{2},\frac{1}{2} \rangle\\
|-,\frac{1}{2},-\frac{1}{2} \rangle \otimes |+,\frac{1}{2},-\frac{1}{2} \rangle%
\end{array}
\right)  ,
\label{TPAMB}
\end{equation}
with
\begin{equation}
M_{A}=\begin{pmatrix} M&0\\0&M \end{pmatrix}, \qquad
M=\left(
\begin{array}
[c]{cccc}%
0 & \frac{1}{\sqrt{2}} & -\frac{1}{\sqrt{2}} & 0\\
1 & 0 & 0 & 0\\
0 & \frac{1}{\sqrt{2}} & \frac{1}{\sqrt{2}} & 0\\
0 & 0 & 0 & 1
\end{array}
\right) .
\label{MTPAM}
\end{equation}

We will denote this basis as the angular momentum (A) basis. Operators in this basis are represented by matrices related 
to those in the TP basis as $O^{A}=M_{A}O^{TP}M^{\dagger}_{A}$. In particular, parity and the generators of 
$L^{\uparrow}_{+}$ take the form
\begin{align}
\Pi^{A}(k)&=\begin{pmatrix}0& \mathbbm{1} \\ \mathbbm{1} & 0 \end{pmatrix}, \qquad
\bm{J}^{A}_{V}=\begin{pmatrix} \bm{J}^{A}_{RL} &0 \\ 0 & \bm{J}^{A}_{RL} \end{pmatrix} \qquad
\bm{K}^{A}_{V}=\begin{pmatrix} \bm{K}^{A}_{RL} &0 \\ 0 & - \bm{K}^{A}_{RL} \end{pmatrix} 
\end{align}
with
\begin{align}
\bm{J}^{A}_{RL} =\left(
\begin{array}
[c]{cc}%
\mathbf{0}_{1\times1} & \bm{0}_{1\times3}\\
\mathbf{0}_{3\times1} & \bm{J}^{(1)}%
\end{array}
\right)  , \qquad 
i\bm{K}^{A}_{RL} =
\left(
\begin{array}
[c]{cc}%
\mathbf{0}_{1\times1} & \bm{V}^{\dagger}\\
\bm{V} & \mathbf{0}_{3\times3}%
\end{array}
\right) 
\label{PiJKA}
\end{align}
where
\begin{equation}
V_{1}^{\dagger}=\frac{1}{\sqrt{2}}(1,0,-1),\qquad V_{2}^{\dagger}=\frac{i}{\sqrt{2}}(1,0,1),
\qquad V_{3}^{\dagger}=(0,-1,0),
\label{Vdagger}
\end{equation}
and $\bm{J}^{(1)}$ denote the angular momentum matrices for $j=1$ in Eq. (\ref{J1}). 

The states on the left of Eq.(\ref{TPAMB}) have well defined $W^{2}$ but not well defined parity. The combinations of well defined 
parity (the Poincar\'{e} states) in the rest frame $|k,m^{2}; j,\lambda, \pi \rangle $ are obtained with the following transformation
\begin{equation}
\left(
\begin{array}
[c]{c}%
|k,m^{2}; 0,0, + \rangle \\
|k,m^{2}; 1,1,- \rangle\\
|k,m^{2}; 1,0,- \rangle \\
|k,m^{2}; 1,-1,- \rangle \\
|k,m^{2}; 0,0,-\rangle \\
|k,m^{2}; 1,1, + \rangle \\
|k,m^{2}; 1,0, +\rangle \\
|k,m^{2}; 1,-1, + \rangle
\end{array}
\right)
= M_{P} 
\left(
\begin{array}
[c]{c}%
|\frac{1}{2},\frac{1}{2}; 0,0\rangle_{RL}\\
|\frac{1}{2},\frac{1}{2}; 1,1\rangle_{RL}\\
|\frac{1}{2},\frac{1}{2}; 1,0\rangle_{RL}\\
|\frac{1}{2},\frac{1}{2}; 1,-1\rangle_{RL}\\
|\frac{1}{2},\frac{1}{2}; 0,0\rangle_{LR}\\
|\frac{1}{2},\frac{1}{2}; 1,1\rangle_{LR}\\
|\frac{1}{2},\frac{1}{2}; 1,0\rangle_{LR}\\
|\frac{1}{2},\frac{1}{2}; 1,-1\rangle_{LR}
\end{array}
\right)
\end{equation}
with 
\begin{equation}
M_{P}= \frac{1}{\sqrt{2}} \begin{pmatrix} \mathbbm{1}_{4} & G \\ \mathbbm{1}_{4} & - G \end{pmatrix}, \qquad
G=  \frac{1}{\sqrt{2}} \begin{pmatrix} 1& 0_{1\times 3} \\ 0_{3\times 1} & -\mathbbm{1}_{3}  \end{pmatrix}.
\end{equation}
The states $\{ |k,m^{2}; j,\lambda, \pi \rangle \}$ form the Poicar\'{e} basis of $V$  and the 
representation of operators in this basis are related to those in the angular momentum basis as $O^{P}=M_{P}O^{A}M^{\dagger}_{P}$.
Parity is diagonal in this basis 
\begin{equation}
\Pi^{P}(k)=Diag (1,-1-1-1,-1,1,1,1)
\end{equation}
and the generators take the form
\begin{equation}
\bm{J}^{P}_{V}=\begin{pmatrix} \bm{J}^{A}_{RL} &0 \\ 0 & \bm{J}^{A}_{RL} \end{pmatrix} \qquad
\bm{K}^{P}_{V}=\begin{pmatrix} \bm{K}^{A}_{RL} &0 \\ 0 &  \bm{K}^{A}_{RL} \end{pmatrix} 
\end{equation}

Notice that in the Poincar\'{e} basis both, parity and $L^{\uparrow}_{+}$ transformations, are block diagonal, thus we obtain a direct sum 
of subspaces irreducible under both, $L^{\uparrow}_{+}$ and parity. Each subspace contains two Poincar\'{e} sectors, $j=0$ and $j=1$ with 
opposite parity. States in these subspaces have the same transformation properties 
under $L^{\uparrow}_{+}$ but have different parity content. In summary
\begin{equation}
\left[\left(\frac{1}{2},0\right)\otimes \left(0,\frac{1}{2}\right)\right] \oplus \left[\left(0,\frac{1}{2}\right) \otimes \left(\frac{1}{2},0\right)\right] = 
\left(\frac{1}{2},\frac{1}{2}\right)_{+-}\oplus  \left(\frac{1}{2},\frac{1}{2}\right)_{-+},
\end{equation} 
where the suffix $+-$ and $-+$ denote the parity content for the corresponding $j=0$ and $j=1$ Poincar\'{e} irreps. It is clear that 
we can work independently with the $(1/2,1/2)_{+-}$ or with the $(1/2,1/2)_{-+}$. 

In the following we will work in the $(1/2,1/2)_{+-}$ representation space, and will denote the generators as 
$\bm{J}^{A}_{RL}\equiv \bm{J}^{P}$,  $\bm{K}^{A}_{RL}\equiv \bm{K}^{P}$ to 
emphasize that these are the representations of the generators in the Poincar\'{e} basis for the subspace $(1/2,1/2)_{+-}$. 
With this convention, the Poincar\'{e} basis for the $(1/2,1/2)_{+-}$ subspace is given by 
$\{ |k,m^{2}; 0,0, + \rangle , |k,m^{2}; 1,\lambda,- \rangle \}$, parity is diagonal $\Pi^{P}(k)=Diag(1,-1,-1,-1)$ and states in this subspace 
transform with the generators
\begin{align}
\bm{J}^{P} =\left(
\begin{array}
[c]{cc}%
\mathbf{0}_{1\times1} & \bm{0}_{1\times3}\\
\mathbf{0}_{3\times1} & \bm{J}^{(1)}%
\end{array}
\right)  , \qquad 
i\bm{K}^{P} =
\left(
\begin{array}
[c]{cc}%
\mathbf{0}_{1\times1} & \bm{V}^{\dagger}\\
\bm{V} & \mathbf{0}_{3\times3}%
\end{array}
\right) 
\label{JKP}
\end{align}
where $\bm{J}^{(1)}$ are the angular momentum matrices for $j=1$ in Eq. (\ref{J1}) and $\bm{V}^{\dagger}$ are the row vectors in 
Eq. (\ref{Vdagger}). 

The representation of the states in the basis $\{ |k,m^{2},j,\lambda,\pi \rangle \}$ is canonical in this basis
\begin{align}
 \phi(k,m^{2},0,0,+)&=\left(
\begin{array}
[c]{c}%
1\\
0\\
0\\
0
\end{array}
\right)  ,\qquad 
\phi(k,m^{2},1,1,-)=\left(
\begin{array}
[c]{c}%
0\\
1\\
0\\
0
\end{array}
\right) , \nonumber \\
 \phi(k,m,1,0,-)&=\left(
\begin{array}
[c]{c}%
0\\
0\\
1\\
0
\end{array}
\right)  ,\qquad 
\phi(k,m^{2},1,-1,-)=\left(
\begin{array}
[c]{c}%
0\\
0\\
0\\
1
\end{array}
\right)  .
\end{align}
These states satisfy the eigenvalue equation
\begin{equation}
\Pi^{P} (k) \, \phi(k,m^{2},j,\lambda,\pi)=\pi \, \phi(k,m^{2},j,\lambda,\pi).
\end{equation}

The boost operator in the Poincar\'{e} basis reads
\begin{equation}
U^{P}(L(p),0)=\frac{1}{m}\left(
\begin{array}
[c]{cccc}%
E & -\frac{p_{+}}{\sqrt{2}}  & p_{3} & \frac{p_{-}}{\sqrt{2}} \\
-\frac{p_{-}}{\sqrt{2}} & \frac{(E+m)^{2}-p_{3}^{2}}{2(E+m)} & -\frac{p_{3} p_{-}}{\sqrt{2}(E+m)} & -\frac{p_{-}^{2}}{2(E+m)}\\
p_{3} & -\frac{p_{3} p_{+}}{\sqrt{2}(E+m)} & \frac{(E+m)^{2}+p_{3}^{2}-p_{+}~p_{-}}{2(E+m)} & \frac{p_{3}~p_{-}}{\sqrt{2}(E+m)}\\
 \frac{p_{+}}{\sqrt{2}} & -\frac{p_{+}^{2}}{2(E+m)} & \frac{p_{3}p_{+}}{\sqrt{2}(E+m)} & \frac{(E+m)^{2}-p_{3}^{2}}{2(E+m)}%
\end{array}
\right),
\label{BP}
\end{equation}
and acting with the 
boost operator on the rest frame states we obtain the following four-components column vectors $\phi_{\lambda\pi}(p,m^{2},j)$
\begin{align}
\phi_{0+}(p,m^{2},0)&=\frac{1}{m}\left(
\begin{array}
[c]{c}%
E\\
-\frac{p_{-}}{\sqrt{2}}\\
p_{3}\\
\frac{p_{+}}{\sqrt{2}}%
\end{array}
\right) ,\qquad 
\phi_{1-}(p,m^{2},1)=\frac{1}{m}\left(
\begin{array}
[c]{c}%
-\frac{p_{+}}{\sqrt{2}}\\
\frac{(E+m)^{2}-p_{3}^{2}}{2(E+m)}\\
-\frac{p_{3}~p_{+}}{\sqrt{2}(E+m)}\\
-\frac{p_{+}^{2}}{2(E+m)}%
\end{array}
\right), \nonumber \\
\phi_{0-}(p,m^{2},1)&=\frac{1}{m}\left(
\begin{array}
[c]{c}%
p_{3}\\
-\frac{p_{3}~p_{-}}{\sqrt{2}(E+m)}\\
\frac{(E+m)^{2}+p_{3}^{2}-p_{+}p_{-}}{2(E+m)}\\
\frac{p_{3}~p_{+}}{\sqrt{2}(E+m)}%
\end{array}
\right), \qquad 
\phi_{-1-}(p,m^{2},1)=\frac{1}{m}\left(
\begin{array}
[c]{c}%
\frac{p_{-}}{\sqrt{2}}\\
-\frac{p_{-}^{2}}{2(E+m)}\\
\frac{p_{3}~p_{-}}{\sqrt{2}(E+m)}\\
\frac{(E+m)^{2}-p_{3}^{2}}{2(E+m)}%
\end{array}\right) .
 \label{soltamb}%
\end{align}
These states satisfy the boosted conditions
\begin{equation}
U^{P}(L(p),0) \Pi^{P} (k) (U^{P}(L(p),0))^{-1} \, \phi_{\lambda\pi}(p,m^{2},j)=\pi \, \phi_{\lambda\pi}(p,m^{2},j).
\end{equation}

A straightforward calculation using Eq. (\ref{BP}) yields
\begin{equation}
U^{P}(L(p),0) \Pi^{P} (k) (U^{P}(L(p),0))^{-1}=\frac{S^{P}_{\mu\nu}p^{\mu}p^{\nu}}{m^{2}}\equiv \frac{S^{P}(p)}{m^{2}},
\label{boostedPi}
\end{equation} 
where the matrices $S^{P}_{\mu\nu}$ satisfy $S^{P}_{\mu\nu}=S^{P}_{\nu\mu}$ and are given by
\begin{align}
S^{P}_{00}&=\left(
\begin{array}{cccc}
 1 & 0 & 0 & 0 \\
 0 & -1 & 0 & 0 \\
 0 & 0 & -1 & 0 \\
 0 & 0 & 0 & -1 \\
\end{array}
\right),\quad
S^{P}_{01}=\left(
\begin{array}{cccc}
 0 & \frac{1}{\sqrt{2}} & 0 & -\frac{1}{\sqrt{2}} \\
 -\frac{1}{\sqrt{2}} & 0 & 0 & 0 \\
 0 & 0 & 0 & 0 \\
 \frac{1}{\sqrt{2}} & 0 & 0 & 0 \\
\end{array}
\right),\quad
S^{P}_{02}=\left(
\begin{array}{cccc}
 0 & \frac{i}{\sqrt{2}} & 0 & \frac{i}{\sqrt{2}} \\
 \frac{i}{\sqrt{2}} & 0 & 0 & 0 \\
 0 & 0 & 0 & 0 \\
 \frac{i}{\sqrt{2}} & 0 & 0 & 0 \\
\end{array}
\right), \nonumber \\
S^{P}_{03}&= \left(
\begin{array}{cccc}
 0 & 0 & -1 & 0 \\
 0 & 0 & 0 & 0 \\
 1 & 0 & 0 & 0 \\
 0 & 0 & 0 & 0 \\
\end{array}
\right), \quad
S^{P}_{11}=\left(
\begin{array}{cccc}
 1 & 0 & 0 & 0 \\
 0 & 0 & 0 & 1 \\
 0 & 0 & 1 & 0 \\
 0 & 1 & 0 & 0 \\
\end{array}
\right),\quad
S^{P}_{12}=\left(
\begin{array}{cccc}
 0 & 0 & 0 & 0 \\
 0 & 0 & 0 & -i \\
 0 & 0 & 0 & 0 \\
 0 & i & 0 & 0 \\
\end{array}
\right) ,\quad
S^{P}_{13}= \left(
\begin{array}{cccc}
 0 & 0 & 0 & 0 \\
 0 & 0 & \frac{1}{\sqrt{2}} & 0 \\
 0 & \frac{1}{\sqrt{2}} & 0 & -\frac{1}{\sqrt{2}} \\
 0 & 0 & -\frac{1}{\sqrt{2}} & 0 \\
\end{array}
\right), \nonumber \\
S^{P}_{22}&= \left(
\begin{array}{cccc}
 1 & 0 & 0 & 0 \\
 0 & 0 & 0 & -1 \\
 0 & 0 & 1 & 0 \\
 0 & -1 & 0 & 0 \\
\end{array}
\right),\quad
S^{P}_{23}= \left(
\begin{array}{cccc}
 0 & 0 & 0 & 0 \\
 0 & 0 & -\frac{i}{\sqrt{2}} & 0 \\
 0 & \frac{i}{\sqrt{2}} & 0 & \frac{i}{\sqrt{2}} \\
 0 & 0 & -\frac{i}{\sqrt{2}} & 0 \\
\end{array}
\right),\quad
S^{P}_{33}= \left(
\begin{array}{cccc}
 1 & 0 & 0 & 0 \\
 0 & 1 & 0 & 0 \\
 0 & 0 & -1 & 0 \\
 0 & 0 & 0 & 1 \\
\end{array}
\right).
\label{SP}
\end{align}
In this case the symmetric tensor satisfy
\begin{equation}
(S^{P})^{\mu}_{~\mu}=-2\mathbbm{1},
\end{equation}
thus we have actually only nine independent matrices in $S^{P}_{\mu\nu}$. 

The four-component vectors $\phi_{\lambda\pi}(p,m^{2},j)$ satisfy the condition
\begin{equation}
[S^{P}_{\mu\nu}p^{\mu}p^{\nu}-\pi m^{2}]\phi_{\lambda\pi}(p,m^{2},1)=0,
\end{equation}
and we can be tempted to consider this condition as the equation of motion in momentum space, i.e. to consider the states 
$\psi (x)=e^{-ip\cdot x} \phi_{\lambda\pi}(p,m^{2},j)$ which satisfy the equation
\begin{equation}
[S^{P}_{\mu\nu}\partial^{\mu}\partial^{\nu}+\pi m^{2}]\psi (x)=0.
\label{pareqv}
\end{equation}
This equation was obtained firstly in Ref. (\cite{Ahluwalia:2000pj}) by a different procedure which yields different representation 
for $S^{\mu\nu}$. 

Actually there is a problem with this equation. A straightforward calculation shows that the following relation holds
\begin{equation}
(S^{P}(p))^{2}=p^{4} \mathbbm{1},
\end{equation}
which can be used when multiplying Eq.(\ref{pareqv}) on the left by $(S^{P}(\partial)-m^{2})$ to show that 
\begin{equation}
(\partial^{4}-m^{4})\psi(x)=0,
\end{equation}
thus this equation suffers of the same problem as the Weinberg equation for the $(1,0)\oplus (0,1)$ representation, i.e., it has (tachyionic) 
solutions with $p^{2}=-m^{2}$ in addition to the $p^{2}=m^{2}$ solutions in Eq. (\ref{soltamb}).

The way out this problem, as we know from our experience with the $(1,0)\oplus (0,1)$ representation, is to project onto 
well defined parity and $p^{2}$ in the $(1/2,1/2)_{+-}$ representation space. 
The projectors onto well defined parity ($\pi=\pm$) subspaces, in the rest frame, 
read%
\begin{equation}
\mathbb{P}_{\pi}(k)=\frac{1}{2}(\mathbbm{1}+\pi\Pi (k)). 
\end{equation}
Boosting the rest frame parity projectors operators we get
\begin{equation}
\mathbb{P}_{\pi}(p)=B(p)\mathbb{P}_{\pi}(k)B^{-1}(p)
=\frac{1}{2}\left( \mathbbm{1}+\pi B(p)\Pi (k)B^{-1}(p)\right).
\end{equation}

In the Poincar\'{e} basis we get 
\begin{equation}
\mathbb{P}^{P}_{\pi}(p)=\frac{1}{2}\left(\mathbbm{1}+\pi \frac{S^{P}(p)}{p^{2}}\right).
\label{PAp}
\end{equation} 
These operators satisfy the projector relations 
\begin{equation}
(\mathbb{P}^{P}_{\pm}(p))^{2}=\mathbb{P}^{P}_{\pm}(p),\qquad 
\mathbb{P}^{P}_{+}(p)\mathbb{P}^{P}_{-}(p)=\mathbb{P}^{P}_{-}(p)\mathbb{P}^{P}_{+}(p)=0,\qquad 
\mathbb{P}^{P}_{+}(p)+\mathbb{P}^{P}_{-}(p)=\mathbbm{1}. 
\end{equation}

States with well defined parity and $p^{2}$ satisfy the following condition
\begin{equation}
 \frac{p^{2}}{m^{2}}\mathbb{P}^{P}_{\pi} (p) \phi_{\lambda\pi}(p,m^{2},1)=\phi_{\lambda\pi}(p,m^{2},1).
\label{eomv}
\end{equation}
It was shown above that for the $(\frac{1}{2}, \frac{1}{2})_{+-}$ representation the eigenvalues of $\Pi$ and $W^{2}$ are correlated, 
states with $\pi=+1$ have $j=0$ and states with $\pi=-1$ have $j=1$ thus it is not necessary to further project onto $W^{2}$ 
eigensubspaces. Notice also that, for a given parity, the solution is constrained to vanish under the action of the projector 
on the opposite parity subspaces, e.g.  the negative parity states satisfy the condition
\begin{equation}
 [p^{2}\mathbb{P}^{P}_{-} (p) -m^{2}]\phi_{\lambda-}(p,m^{2},1)=0,
\label{eomvp}
\end{equation}
and acting on the left with $p^{2} \mathbb{P}^{P}_{+} (p) $ we get the constraint
\begin{equation}
p^{2} \mathbb{P}^{P}_{+} (p) \phi_{\lambda-}(p,m^{2},1)=0.
\end{equation}

In configuration space we can define the negative parity states as $V(x)=e^{-ip\cdot x}\phi_{\lambda-}(p,m^{2},1)$. 
These states satisfy the equation
\begin{equation}
\left( \partial^{2}  \mathbb{P}^{P}_{-}(\partial)+ m^{2} \right) V(x)=0,
\label{eomv0}
\end{equation}
This equation can be rewritten as
\begin{equation}
\left( \Sigma^{P}_{\mu\nu}\partial^{\mu}\partial^{\nu} + m^{2} \right) V(x)=0,
 \label{eomv}
\end{equation}
with 
\begin{equation}
 \Sigma^{P}_{\mu\nu}=\frac{1}{2}(g_{\mu\nu} - S^{P}_{\mu\nu}).
\end{equation}

Applying on the left Eq. of (\ref{eomv0}) the operator $\partial^{2}\mathbb{P}^{P}_{+}(\partial)+m^{2}$ we obtain
\begin{equation}
(\partial^{2}\mathbb{P}^{P}_{+}(\partial)+m^{2})(\partial^{2}\mathbb{P}^{P}_{-}(\partial)+m^{2})V(x)=m^{2}(\partial^{2}+m^{2})V(x)=0,
\end{equation}
thus the four components of the vector $V(x)$ satisfies indeed the Klein-Gordon equation. Also, applying the operator 
$\partial^{2}\mathbb{P}^{P}_{+}(\partial)$ we 
obtain the constraint
\begin{equation}
\left[\partial^{2}  \mathbb{P}^{P}_{+}(\partial) \right] V(x)=0,
\end{equation}
which can be rewritten as
\begin{equation}
\left( R^{P}_{\mu\nu}\partial^{\mu}\partial^{\nu}  \right) V(x)=0,
\end{equation}
with 
\begin{equation}
R^{P}_{\mu\nu}=\frac{1}{2}(g_{\mu\nu} + S^{P}_{\mu\nu}).
\end{equation}
This condition eliminates the $j=0$ and positive parity component of the general field described by the four-component field $V(x)$.

The equation of motion ($\ref{eomv}$) can be obtained from the following Lagrangian
\begin{equation}
{\cal{L}}=- \partial^{\mu}\bar{V}(x)\Sigma^{P}_{\mu\nu}\partial^{\nu}V(x) + m^{2}\bar{V}(x)V(x),
\end{equation}
where $\bar{V}(x)\equiv V^{\dagger}(x)S^{P}_{00}=V^{\dagger}(x)\Pi^{P}$. The matrices $S^{P}_{\mu\nu}$ satisfy
\begin{equation}
\Pi^{P}S^{P\dagger}_{\mu\nu}\Pi^{P}=S^{P}_{\mu\nu},
\end{equation}
which yields a Hermitian Lagrangian. The propagator is obtained as
\begin{equation}
iS^{P}_{V}(p)=\frac{p^{2}\mathbb{P}^{P}_{+}(p)-m^{2}}{m^{2}(p^{2}-m^{2}+i\epsilon)}
=\frac{-\mathbb{P}^{P}_{-}(p) + \frac{p^{2} - m^{2}}{m^{2}}\mathbb{P}^{P}_{+}(p)}{p^{2}-m^{2}+i\epsilon}.
\end{equation}
Notice the close resemblance of this equation with the propagator for the $(1,0)\oplus (0,1)$ field in Eq. (\ref{proptensor}). In both 
cases the bad behavior at high energies stems from the term proportional to $p^{2}/m^{2}$ in the numerator. Fortunately, in the 
development of the standard model we have learnt how to solve this problem in the case of fields transforming in the $(1/2,1/2)$ 
representation. The solution is given by the Higgs mechanism for the generation of mass for the gauge bosons where the starting 
point is the theory for a massless gauge boson. 

\subsection{Massless limit, parity and emerging symmetry}

Our derivation has been done assuming $m\neq 0$. A similar treatment for the massless case would require to work out 
the construction from first principles of the corresponding little group of the Poincar\'e group, in this case the $ISO(2)$ group. 
Notice however that both the equation of motion and the Lagrangian have a soft $m\to 0$ limit. In the analysis of this limit 
it is convenient to rewrite the Lagrangian in terms of the parity projectors
\begin{align}
{\cal{L}}&= - \partial^{\mu}\left(\bar{V}(x)\Sigma^{P}_{\mu\nu}\partial^{\nu}V(x)\right) 
+ \bar{V}(x)\Sigma^{P}_{\mu\nu}\partial^{\mu}\partial^{\nu}V(x) + m^{2}\bar{V}(x)V(x) \nonumber\\
&=  \partial^{\mu}F_{\mu}(x) +\bar{V}(x)[\partial^{2}\mathbb{P}^{P}_{-}(\partial) + m^{2}] V(x),
\end{align}
where $F_{\mu}(x)= - \bar{V}(x)\Sigma^{P}_{\mu\nu}\partial^{\nu}V(x)$. 

Consider now the massless limit. The Lagrangian in this case reads
\begin{equation}
{\cal{L}}=- \partial^{\mu}\bar{V}(x)\Sigma^{P}_{\mu\nu}\partial^{\nu}V(x)=   
\partial^{\mu}F_{\mu}(x) +\bar{V}(x)\partial^{2}\mathbb{P}^{P}_{-}(\partial)  V(x),
\end{equation} 
the equation of motion reduces to
\begin{equation}
 \partial^{2}  \mathbb{P}^{P}_{-}(\partial)V(x)=0,
\label{eomv0m0}
\end{equation}
and in this limit we loose the constraint which eliminates the unphysical parity component. Furthermore, the operator in the kinetic term is a projector 
(modulo a surface term) and the propagator is ill-defined. However, it is clear that if $V(x)$ satisfies Eq.(\ref{eomv0m0}), the field
$V^{\prime}(x)=V(x)+\partial^{2} \mathbb{P}^{P}_{+}(\partial)W(x)$ with $W(x)$ an arbitrary field transforming in the $(1/2,1/2)_{+-}$ 
representation also satisfies this equation. At the Lagrangian level
the change $V(x)\to V(x)+\delta V(x)$ with $\delta V(x)=\partial^{2} \mathbb{P}^{P}_{+}(\partial)W(x)$ induces the following change 
\begin{align}
\delta{\cal{L}}&=  \partial^{\mu}(\delta F_{\mu}(x))  + \delta\bar{V}(x) [\partial^{2}\mathbb{P}^{P}_{-}(\partial) ] V(x)
+ \bar{V}(x) [\partial^{2}\mathbb{P}^{P}_{-}(\partial) ] \delta V(x) \nonumber \\
&=  \partial^{\mu}(\delta F_{\mu}(x)) +   \partial^{2} \bar{W}(x) \mathbb{P}^{P}_{+}(\partial) \partial^{2}\mathbb{P}^{P}_{-}(\partial) V(x) 
+\bar{V}(x)\partial^{2}\mathbb{P}^{P}_{-}(\partial) \partial^{2}\mathbb{P}^{P}_{+}(\partial) W(x)  =  \partial^{\mu}(\delta F_{\mu}(x)) ,
\end{align}
where we used $ \mathbb{P}^{P}_{+}(\partial)\mathbb{P}^{P}_{-}(\partial)= \mathbb{P}^{P}_{-}(\partial)\mathbb{P}^{P}_{+}(\partial)=0 $. 
We conclude that in the massless case the 
action is invariant under arbitrary changes in the unphysical positive parity component of $V(x)$. 

In summary, in the massless limit we loose the constraint which eliminates the undesired (positive) parity component, but at the same time 
a symmetry emerges which makes this component arbitrary, thus this component must still be unphysical. We can enforce the vanishing 
of the unphysical parity component incorporating the corresponding constraint in the Lagrangian and working instead with the 
following Lagrangian
\begin{align}
{\cal{L}}&=- \partial^{\mu}\bar{V}(x)\Sigma^{p}_{\mu\nu}\partial^{\nu}V(x) 
- \frac{1}{\xi}\partial^{\mu}\bar{V}(x)R^{P}_{\mu\nu}\partial^{\nu}V(x) \\
&= - \partial^{\mu}\left(\bar{V}(x)\Sigma^{P}_{\mu\nu}+\frac{1}{\xi}R^{P}_{\mu\nu}) \partial^{\nu}V(x)\right) 
 +\bar{V}(x)\partial^{2}[\mathbb{P}^{P}_{-}(\partial) +\frac{1}{\xi}\mathbb{P}^{P}_{+}(\partial) ] V(x),
\label{Lagxi}
\end{align}
where $\xi$ is an arbitrary parameter. The equation of motion with this new term is
\begin{equation}
\partial^{2}[\mathbb{P}^{P}_{-}(\partial) +\frac{1}{\xi}\mathbb{P}^{P}_{+}(\partial) ] V(x)=0,
\label{eomvoxi}
\end{equation}
and acting on the left of this equation with $\mathbb{P}^{P}_{+}(\partial) $ we get 
\begin{equation}
\frac{1}{\xi}\partial^{2}\mathbb{P}^{P}_{+}(\partial)  V(x)=0,
\end{equation}
recovering the constraint for finite $\xi$ and using it in Eq. (\ref{eomvoxi}) we obtain Eq. (\ref{eomv0m0}) for the physical degrees of freedom.
The kinetic energy operator is invertible now and we get the propagator
\begin{equation}
iS^{P}_{V}(p,\xi)=\frac{-\mathbb{P}^{P}_{-}(p)-\xi \mathbb{P}^{P}_{+}(p)}{(p^{2}+i\epsilon)}
=\frac{-\mathbbm{1}+(1-\xi) \mathbb{P}^{P}_{+}(p)}{p^{2}+i\epsilon}.
\label{propva}
\end{equation}
The arbitrariness of $\xi$ reflects the unphysical nature of the opposite parity components and results for scattering amplitudes 
cannot depend on this parameter. It is clear that the numerator in the propagator is ${\cal{O}}(1)$ and this propagator is well behaved 
at high energies for every value of $\xi$ although it takes its simpler form for $\xi=1$.

\subsection{"Minkowskian" basis }
We can recognize this structure if we work in a different basis for the $(1/2,1/2)_{+-}$ representation. So far, in our construction  $V(x)$ is 
a four-component column vector, the four components being the coefficients of the expansion of the state in the chosen basis and 
transforming under boosts with $U^{P}(L(p),0)$ in Eq.(\ref{BP}).

Let us now change to the "Minkowskian" (M) basis related to the Poincar\'{e} basis as 
\begin{equation}
\left(
\begin{array}
[c]{c}%
|\varepsilon^{(0)}(k)\rangle\\
|\varepsilon^{(1)}(k)\rangle\\
|\varepsilon^{(2)}(k)\rangle\\
|\varepsilon^{(3)}(k)\rangle
\end{array}
\right)  =\left(
\begin{array}
[c]{cccc}%
1 & 0 & 0 & 0\\
0 & -\frac{1}{\sqrt{2}} & 0 & \frac{1}{\sqrt{2}}\\
0 & -\frac{i}{\sqrt{2}} & 0 &- \frac{i}{\sqrt{2}}\\
0 & 0 & 1 & 0
\end{array}
\right)  
\left(
\begin{array}
[c]{c}%
|k,m^{2},0,0,+\rangle\\
|k,m^{2},1,1,-\rangle\\
|k,m^{2},1,0,-\rangle\\
|k,m^{2},1,-1,-\rangle
\end{array}
\right)  .
\label{PBtoMB}
\end{equation}
In this basis the parity operator remains diagonal 
\begin{equation}
\Pi^{M}=Diag(1,-1-1,-1),
\end{equation}
thus $|\varepsilon^{(0)}(k)\rangle$ is a state with positive parity and $|\varepsilon^{(i)}(k)\rangle$, $i=1,2,3$ have negative parity. 
These states are all eigenstates of $W^{2}$ but only $|\varepsilon^{(0)}\rangle$ and $|\varepsilon^{(3)}\rangle$ are eigenstates 
of $J_{3}$. The generators of $L^{\uparrow}_{+}$ in the new basis read %
\begin{equation}
J_{1}^{M}=\left[
\begin{array}
[c]{cccc}%
0 & 0 & 0 & 0\\
0 & 0 & 0 & 0\\
0 & 0 & 0 & -i\\
0 & 0 & i & 0
\end{array}
\right] , ~~~
J_{2}^{M}=\left[  {%
\begin{array}
[c]{rcrc}%
0 & 0 & 0 & 0\\
0 & 0 & 0 & i\\
0 & 0 & 0 & 0\\
0 & -i & 0 & 0
\end{array}
}\right] , ~~~
J_{3}^{M}=\left[  {%
\begin{array}
[c]{rccr}%
0 & 0 & 0 & 0\\
0 & 0 & -i & 0\\
0 & i & 0 & 0\\
0 & 0 & 0 & 0
\end{array}
}\right] ,
\end{equation}

\begin{equation}
K_{1}^{M}=i\left[  {%
\begin{array}
[c]{rrrr}%
0 & 1 & 0 & 0\\
1 & 0 & 0 & 0\\
0 & 0 & 0 & 0\\
0 & 0 & 0 & 0
\end{array}
}\right],  ~~~
K_{2}^{M}=i\left[  {%
\begin{array}
[c]{rrrr}%
0 & 0 & 1 & 0\\
0 & 0 & 0 & 0\\
1 & 0 & 0 & 0\\
0 & 0 & 0 & 0
\end{array}
}\right] , ~~~
K_{3}^{M}=i\left[  {%
\begin{array}
[c]{rrrr}%
0 & 0 & 0 & 1\\
0 & 0 & 0 & 0\\
0 & 0 & 0 & 0\\
1 & 0 & 0 & 0
\end{array}
}\right] . 
\end{equation}
These are exactly the matrix representations for the generators in the classical Minkowski space, which can be covariantly 
written as the components of a second rank tensor, $J_{V}^{0i}=K^{M}_{i}$, $ J_{V}^{ij}=\epsilon^{ijk} J^{M}_{k}$, where
\begin{equation}
(J_{V}^{\mu\nu})^{\alpha}_{~\beta}=i(g^{\mu\alpha}g^{\nu}_{\beta} - g^{\mu}_{\beta}g^{\nu\alpha}). 
\label{JV}
\end{equation}

The four components of the representation of each of the states in the basis, 
$\{ |\varepsilon^{(0)}(k) \rangle,|\varepsilon^{(1)}(k)\rangle,|\varepsilon^{(2)}(k)\rangle, |\varepsilon^{(3)}(k)\rangle \}$, 
transform under $L^{\uparrow}_{+}$ with
\begin{equation}
U^{M}(\Lambda,0)^{\alpha}_{~\beta}= \left(e^{-i(\bm{J}^{M}\cdot \bm{\theta} + \bm{K}^{M}\cdot \bm{\phi} )}\right)^{\alpha}_{\quad\beta}
=\Lambda^{\alpha}_{~\beta}
\end{equation}
i.e., as the components of a four-vector in the classical Minkowski space. 
The explicit form of the boost operator in this basis is  
\begin{equation}
U^{M}(L(p),0)=\left(
\begin{array}
[c]{cccc}%
\frac{E}{m} & \frac{p_{x}}{m} & \frac{p_{y}}{m} & \frac{p_{z}}{m}\\
\frac{p_{x}}{m} & 1+\frac{p_{x}^{2}}{m(E+m)} & \frac{p_{x}p_{y}}{m(E+m)} &
\frac{p_{x}p_{z}}{m(E+m)}\\
\frac{p_{y}}{m} & \frac{p_{x}p_{y}}{m(E+m)} & 1+\frac{p_{y}^{2}}{m(E+m)} &
\frac{p_{y}p_{z}}{m(E+m)}\\
\frac{p_{z}}{m} & \frac{p_{x}p_{z}}{m(E+m)} & \frac{p_{y}p_{z}}{m(E+m)} &
1+\frac{p_{z}^{2}}{m(E+m)}%
\end{array}
\right) .
\label{boostMB}
\end{equation}

In the rest frame and in the "Minkowskian basis" the representation of the states 
$\{ |\varepsilon^{(0)}(k)\rangle,|\varepsilon^{(1)}(k)\rangle,|\varepsilon^{(2)}(k)\rangle, |\varepsilon^{(3)}(k)\rangle \}$ is canonical. 
Denoting the corresponding four-component vectors
as $\{ \varepsilon^{(a)}(k) \}$, $a=0,1,2,3$, their components are given by $(\varepsilon^{(a)} (k))_{\mu}=g^{a}_{\mu}$. For a frame 
where the particle has momentum $p$, using the boost operator in Eq. (\ref{boostMB}) we obtain
\begin{align}
\varepsilon^{(0)}(p)&=\frac{1}{m}\left(
\begin{array}
[c]{c}%
E\\
p_{x}\\
p_{y}\\
p_{z}%
\end{array}
\right),  \qquad 
\varepsilon^{(1)}(p)=\frac{1}{m(E+m)}\left(
\begin{array}
[c]{c}%
(E+m)p_{1}\\
m(E+m)+p_{x}^{2}\\
p_{x}~p_{y}\\
p_{x}~p_{z}%
\end{array}
\right)  \nonumber \\
\varepsilon^{(2)}(p)&=\frac{1}{m(E+m)}\left(
\begin{array}
[c]{c}%
(E+m)p_{y}\\
p_{x}~p_{y}\\
m(E+m)+p_{y}^{2}\\
p_{y}~p_{z}%
\end{array}
\right), \qquad
\varepsilon^{(3)}(p)=\frac{1}{m(E+m)}\left(
\begin{array}
[c]{c}%
(E+m)p_{z}\\
p_{z}~p_{x}\\
p_{z}~p_{y}\\
m(E+m)+p_{z}^{3}%
\end{array}
\right).
\end{align}
Notice that the components of the positive parity state $\varepsilon^{(0)}(p)$ are explicitly covariant 
$(\varepsilon^{(0)}(p))^{\alpha}=\frac{p^{\alpha}}{m}$ 
and it satisfies $p_{\alpha}(\varepsilon^{0}(p))^{\alpha}=m\neq 0$. The components of the remaining (negative parity) states 
can also be written covariantly as
\begin{equation}
(\varepsilon^{(i)}(p))_{\alpha}= L(p)^{i}_{~\alpha},
\end{equation}
and satisfy $p^{\alpha}(\varepsilon^{(i)}(p))_{\alpha}= 0$. We can understand these results if we work out the parity projectors 
and write the Poincar\'{e} and parity projection condition in the Minkowskian basis. The parity projectors in Eq (\ref{PAp}) 
are written in terms of the symmetric tensor matrices in Eq. (\ref{SP}) which in the Minkowskian basis read
\begin{equation}
(S^{M}_{\mu\nu})^{\alpha}_{~\beta}= - g_{\mu\nu}g^{\alpha}_{\beta}+g^{\alpha}_{\mu}g^{\beta}_{\nu} + g^{\alpha}_{\nu}g^{\beta}_{\mu}, 
\end{equation}
and the parity projectors take the form
\begin{align}
\lbrack \mathbb{P}^{M}_{+}(p)]^{\alpha}_{~\beta}  &  =\frac{p^{\alpha}p_{\beta}}{p^{2}}
\label{Proyp}\\
\lbrack \mathbb{P}^{M}_{-}(p)]^{\alpha}_{~\beta}  &  =g^{\alpha}_{~\beta}-\frac{p^{\alpha}p_{\beta}}{p^{2}}.
\end{align}
The negative parity and Poincar\'e projection in Eq. (\ref{eomvp}) can be rewritten as
\begin{equation}
[\frac{p^{2}}{m^{2}}g^{\alpha}_{~\beta}-\frac{p^{\alpha}p_{\beta}}{m^{2}}](\varepsilon^{(i)})^{\beta}(p)=(\varepsilon^{(i)})^{\alpha}(p).
\end{equation}

In terms of $\varepsilon^{(i)}_{\alpha\beta}(p)\equiv p_{\alpha}\varepsilon^{(i)}_{\beta}(p)-p_{\beta}\varepsilon^{(i)}_{\alpha}(p)$ 
it reads
\begin{equation}
p^{\alpha}\varepsilon^{i}_{\alpha\beta}(p)-m^{2}\varepsilon^{(i)}_{\beta}(p)=0.
\end{equation}
States of well defined momentum in configuration space are the plane waves $V^{(i)}_{\alpha}(x)=\varepsilon^{(i)}_{\alpha}(p)e^{-ip.x}$ 
which satisfy
\begin{equation}
\partial^{\alpha}F^{(i)}_{\alpha\beta}(x) + m^{2}V^{(i)}_{\beta}(x)=0,
\end{equation}
where $F^{(i)}_{\alpha\beta}(x)\equiv \partial_{\alpha}V^{(i)}_{\beta}(x)-\partial_{\beta}V^{(i)}_{\alpha}(x)$.

This construction makes clear that the Proca equation is just the covariant projection on the appropriate Poincar\'{e} orbit and on 
the negative parity subspace of the {\it{quantum}} $(\frac{1}{2},\frac{1}{2})_{+-}$ representation of the HLG where states in the rest frame 
carry well defined quantum numbers of the CSCO dictated by the full Poincar\'{e} symmetry, written in a specific (Minkowskian) 
basis for this representation space. 

The symmetric tensor appearing in the Lagrangian in this basis reads 
\begin{align}
(\Sigma_{\mu\nu})_{\alpha\beta}&=  \frac{1}{2}\left[ g_{\mu\nu}g_{\alpha\beta} -(S^{M}_{\mu\nu})_{\alpha\beta}\right]
=g_{\mu\nu}g_{\alpha\beta} -\frac{1}{2}( g_{\alpha\mu}g_{\nu\beta} + g_{\alpha\nu}g_{\mu\beta}), 
\end{align}
and the Lagrangian can be written as
\begin{align}
{\cal{L}}&= - \partial^{\mu}\bar{V}^{\alpha}(x)(\Sigma_{\mu\nu})_{\alpha\beta}\partial^{\nu}V^{\beta}(x) + m^{2}\bar{V}^{\alpha}(x)V_{\alpha}(x)
 \nonumber \\
&=-\frac{1}{2} \bar{V}^{\mu\alpha} V_{\mu\alpha}  + m^{2}\bar{V}^{\alpha}V_{\alpha} 
- \partial^{\mu}\left( \bar{V}^{\alpha}(g_{\mu\alpha}g_{\nu\beta}-g_{\mu\beta}g_{\nu\alpha})\partial^{\nu}V^{\beta} \right),
\end{align}
where $V^{\mu\alpha}=\partial^{\mu}V^{\alpha}-\partial^{\alpha}V^{\mu}$ and modulo a surface term we obtain the conventional Proca 
Lagrangian. The propagator in the Minkowskian basis reads
\begin{equation}
i(S^{M}_{V}(p))^{\alpha}_{~\beta}=\frac{- (\mathbb{P}^{M}_{-}(\mathbf{p}))^{\alpha}_{~\beta} 
+ \frac{p^{2} - m^{2}}{m^{2} } (\mathbb{P}^{M}_{+}(\mathbf{p}))^{\alpha}_{~\beta}}{p^{2}-m^{2}+i\epsilon} 
=\frac{-g^{\alpha}_{~\beta} + \frac{p^{\alpha}p_{\beta}}{m^{2}}}{p^{2}-m^{2}+i\epsilon}.
\end{equation}

\subsection{Space-time nature of gauge invariance}

In the massless case the Lagrangian in the Minkowskian basis can be rewritten as
\begin{align}
{\cal{L}}&= - \partial^{\mu}\bar{V}^{\alpha}(x)( \Sigma_{\mu\nu})_{\alpha\beta} \partial^{\nu}V^{\beta}(x)  
=-\frac{1}{2} \bar{V}^{\mu\alpha} V_{\mu\alpha} 
+ \partial^{\mu}\left( \frac{1}{2}\bar{V}^{\alpha}(g_{\mu\alpha}g_{\nu\beta}-g_{\mu\beta}g_{\nu\alpha})\partial^{\nu}V^{\beta}  \right).
\end{align}
This Lagrangian is invariant under the transformations 
$\delta V(x)=\partial^{2} \mathbb{P}^{A}_{+}(\partial)W(x)$ which in the Minkowskian basis reads
\begin{equation}
\delta V^{\alpha}(x)=\partial^{2} (\mathbb{P}^{M}_{+}(\partial))^{\alpha}_{~\beta}W^{\beta}
= \partial^{\alpha}\partial_{\beta}W^{\beta}(x)= \partial^{\alpha} \theta(x),
\end{equation}
with $\theta(x)$ an arbitray scalar function. This is the conventional gauge freedom of a {\it{classical}} massless vector field and the first 
principles construction done here shows that, {\it in Hilbert space,  gauge transformations are just arbitrary changes in the unphysical 
positive parity component of the field transforming in the $(1/2,1/2)_{+-}$ representation space}. Furthermore, the $\xi$-term included in the 
Lagrangian to ensure that the unphysical parity component of the field vanishes is just the conventional gauge fixing term
modulo a surface term. Indeed, in the Minkowskian basis 
\begin{equation}
(R_{\mu\nu})_{\alpha\beta}= \frac{1}{2}\left[ g_{\mu\nu}g_{\alpha\beta} +(S^{M}_{\mu\nu})_{\alpha\beta}\right]
=\frac{1}{2}( g_{\alpha\mu}g_{\nu\beta} + g_{\alpha\nu}g_{\mu\beta}). 
\end{equation}
and the Lagrangian in Eq. (\ref{Lagxi}) reads
\begin{align}
{\cal{L}}&= - \partial^{\mu}\bar{V}^{\alpha}(x)[( \Sigma_{\mu\nu})_{\alpha\beta} + \frac{1}{\xi}(R_{\mu\nu})_{\alpha\beta}] \partial^{\nu}V^{\beta}(x)  
 \nonumber \\
&=-\frac{1}{2} \bar{V}^{\mu\alpha} V_{\mu\alpha} - \frac{1}{\xi}(\partial^{\alpha}\bar{V}_{\alpha})(\partial^{\beta}V_{\beta})
+ \partial^{\mu}\left(\frac{1}{2}(\frac{1}{\xi}+1) \bar{V}^{\alpha}(g_{\mu\alpha}g_{\nu\beta}-g_{\mu\beta}g_{\nu\alpha})\partial^{\nu}V^{\beta} \right).
\end{align}
Finally, in this basis the propagator in Eq.(\ref{propva}) reads
\begin{equation}
i\left(S^{M}_{V}(p,\xi)\right)^{\alpha}_{~\beta}
=\frac{(-\mathbb{P}^{A}_{-}(p))^{\alpha}_{~\beta}-\xi (\mathbb{P}^{A}_{+}(p))^{\alpha}_{~\beta}}{(p^{2}+i\epsilon)}
=\frac{- g^{\alpha}_{~\beta}+(1-\xi) \frac{p^{\alpha}p_{\beta}}{p^{2}}}{p^{2}+i\epsilon}.
\label{propvm}
\end{equation}

It is worth to remark that it is in the Minkowskian basis, where the covariant properties of the $(1/2,1/2)_{+-}$ field are explicit, that we
can identify the emergent symmetry related to arbitrary changes in the unphysical parity sector as the classical gauge symmetry of the 
massless vector field. The gauge symmetry dictates the interactions in the standard model when we impose it as a local symmetry, 
and this can be done only in this basis. Furthermore, mass terms are generated via the Higgs 
mechanism after the local gauge invariance has been imposed, thus also the Higgs mechanism works only in this basis. 
For a charged scalar field interacting with a massless vector field the action for following Lagrangian
\begin{equation}
{\cal{L}}=(D^{\alpha}\phi)^{\dagger} D_{\alpha}\phi - \mu^{2}\phi^{\dagger}\phi - \lambda \left(\phi^{\dagger}\phi \right)^{2} 
- \partial^{\mu}\bar{V}^{\alpha}(x)\left[( \Sigma_{\mu\nu})_{\alpha\beta} + \frac{1}{\xi}(R_{\mu\nu})_{\alpha\beta}\right] \partial^{\nu}V^{\beta}(x) ,
\end{equation}  
where, in the Minkowskian basis
\begin{equation}
D_{\alpha}\phi=\partial_{\alpha} +i g V_{\alpha},
\end{equation}
is invariant under the following transformations
\begin{align}
\phi &\to e^{-ig \partial^{\beta}W_{\beta}(x) }\phi \\
V^{\alpha} &\to V^{\alpha} + ( \partial^{2} \mathbbm{P}_{+}(\partial))^\alpha_{~\beta}W^{\beta} 
=  V^{\alpha} + \partial^{\alpha} (\partial_{\beta}W^{\beta}(x)),
\end{align}
for arbitrary $W^{\beta}(x)$ transforming in the $(1/2,1/2)_{+-}$ representation. Under spontaneous symmetry breaking ($\mu^{2}<0$) we get 
$\phi=e^{i\chi}(\sigma +v)/\sqrt{2}=(\chi+\sigma+v+...)/\sqrt{2}$ which provides a mass $m=gv$ to the vector field such that the kinetic term
of the vector field after spontaneous symmetry breaking reads
\begin{align}
{\cal{L}}^{V}_{kin}&=- \partial^{\mu}\bar{V}^{\alpha}(x)\left[( \Sigma_{\mu\nu})_{\alpha\beta} 
+ \frac{1}{\xi}(R_{\mu\nu})_{\alpha\beta}\right] \partial^{\nu}V^{\beta}(x) +m^{2}\bar{V}^{\alpha}V_{\alpha} 
= -\partial^{\mu}\left(  \bar{V}^{\alpha}(x)\left[( \Sigma_{\mu\nu})_{\alpha\beta} 
+ \frac{1}{\xi}(R_{\mu\nu})_{\alpha\beta}\right] \partial^{\nu}V^{\beta}(x) \right)  \nonumber \\
&+  \bar{V}^{\alpha}(x)\left[( \partial^{2} \mathbb{P}_{-}(\partial))_{\alpha\beta} 
+ \frac{1}{\xi}(\partial^{2} \mathbb{P}_{-}(\partial))_{\alpha\beta} +m^{2}g_{\alpha\beta} \right] V^{\beta}(x). 
\end{align}
The propagator in this case is given by
 \begin{align}
 iS_{V}= - \frac{ \mathbb{P}_{-}(p)+\frac{\xi (p^{2}-m^{2})}{p^{2}-\xi m^{2}} \mathbb{P}_{+}(p) }{p^{2}-m^{2}+i\epsilon} 
 =- \frac{ \mathbbm{1} + (\xi -1) \frac{p^{2}}{p^{2}-\xi m^{2}} \mathbb{P}_{+}(p) }{p^{2}-m^{2}+i\epsilon}.
 \end{align}
This is the conventional propagator for a vector field in the $\xi$ gauge.

With this insight, we address the massless case of the tensor field and the analogous symmetry related to transformations in the 
unphysical sector. 

\section{Massless limit and gauge invariance for the tensor matter field }

Let us consider now the massless limit for the tensor field and perform the same analysis done for the vector field. We first rewrite the
massless Lagrangian as
\begin{align}
{\cal{L}}&= - \partial^{\mu}\left( \bar{\psi}(x)\Sigma_{\mu\nu}\partial^{\nu}\psi(x) \right) 
+ \bar{\psi}(x)\Sigma_{\mu\nu}\partial^{\mu}\partial^{\nu}\psi(x) \nonumber\\
&=  \partial^{\mu}G_{\mu}(x) +\bar{\psi}(x)[\partial^{2}\mathbb{P}_{+}(\partial) ] \psi(x),
\end{align}
where $G_{\mu}(x)= - \bar{\psi}(x)\Sigma_{\mu\nu}\partial^{\nu}\psi(x)$. This Lagrangian yields the following equation of motion
\begin {equation}
[\partial^{2}\mathbb{P}_{+}(\partial) ] \psi(x)=0.
\end{equation}
Notice that if $\psi(x)$ satisfies this equation so does $\psi^{\prime}(x)=\psi(x)+\partial^{2}\mathbb{P}_{-}(\partial)\chi(x)$ 
where $\chi(x)$ is an arbitrary field transforming in the $(1,0)\oplus (0,1)$ representation. Also in this case,
in the massless limit, we loose the constraint on the unphysical parity component of the field and an emergent symmetry arises 
related to arbitrary changes in the unphysical  part of $\psi$ (negative parity component in this case) 
\begin{equation}
\delta \psi(x)=\partial^{2}\mathbb{P}_{+}(\partial)\chi(x)=\frac{1}{2} (\partial^{2}- S^{\mu\nu}\partial_{\mu}\partial_{\nu})\chi (x) 
\equiv R^{\mu\nu}\partial_{\mu}\partial_{\nu}\chi (x).
\end{equation}
At the Lagrangian level, the change of the Lagrangian is given by
\begin{align}
\delta{\cal{L}}&=  \partial^{\mu}(\delta G_{\mu}(x))  + \delta\bar{\psi}(x) [\partial^{2}\mathbb{P}_{+}(\partial) ] \psi(x)
+ \bar{\psi}(x) [\partial^{2}\mathbb{P}_{+}(\partial) ] \delta \psi(x) \nonumber \\
&=  \partial^{\mu}(\delta G_{\mu}(x)) +   \partial^{2} \bar{\chi}(x) \mathbb{P}_{-}(\partial) \partial^{2}\mathbb{P}_{+}(\partial) \psi(x) 
+\bar{\psi}(x)\partial^{2}\mathbb{P}_{+}(\partial) \partial^{2}\mathbb{P}_{-}(\partial) \chi(x)  =  \partial^{\mu}(\delta G_{\mu}(x)) ,
\end{align}
thus the action is invariant under this transformation.

The kinetic operator in the Lagrangian is not invertible in the massless case (it is a projector). We can solve both problems adding a 
parity-fixing term in the Lagrangian
\begin{align}
{\cal{L}}=&= - \partial^{\mu} \bar{\psi}(x)\Sigma_{\mu\nu}\partial^{\nu}\psi(x) 
- \frac{1}{\xi} \partial^{\mu}\bar{\psi}(x)R_{\mu\nu}\partial^{\nu}\psi(x) \\
&= - \partial^{\mu}\left( \bar{\psi}(x)(\Sigma_{\mu\nu}+\frac{1}{\xi}R_{\mu\nu}) \partial^{\nu}\psi(x) \right) 
 +\bar{\psi}(x)\partial^{2}[\mathbb{P}_{+}(\partial) +\frac{1}{\xi}\mathbb{P}_{-}(\partial) ] \psi(x),
\end{align}
which yields the following equation of motion
\begin {equation}
\partial^{2}[\mathbb{P}_{+}(\partial) +\frac{1}{\xi}\mathbb{P}_{-}(\partial)] \psi(x)=0.
\end{equation}
Acting on the left of this equation with $\mathbb{P}_{-}(\partial)$ we obtain the constraint
\begin {equation}
\frac{1}{\xi}\partial^{2}\mathbb{P}_{-}(\partial) \psi(x)=0,
\end{equation}
which eliminates the unphysical parity component. Also, the kinetic operator is invertible for finite $\xi$ and the propagator is given as
\begin{align}
iS_T(p, \xi)=\frac{\mathbb{P}_{+}(p) + \xi \mathbb{P}_{-}(p)}{p^{2}+i\epsilon}
= \frac{\mathbbm{1} + (\xi-1) \mathbb{P}_{-}(p)}{p^{2}+i\epsilon}.
\end{align}
This propagator is well behaved in the untraviolet and, in the view of the connection of the emergent symmetry with the gauge symmetry 
obtained in the case of the $(1/2,1/2)_{+-}$ representation it is natural to wonder if this connection holds for the $(1,0)\oplus (0,1)$ 
representation. In this concern, we have learnt from the $(1/2,1/2)_{+-}$ case that it is convenient to 
work in the Minkowskian basis. 

It is possible to write the formalism for the $(1,0)\oplus (0,1)$ in a tensor formalism. Indeed, first we perform the following change of basis
\begin{align}
\begin{pmatrix} |+,v_{1}\rangle \\ |+,v_{2} \rangle \\ |+,v_{3} \rangle \\  |-,v_{1}\rangle \\ |-,v_{2}\rangle \\ |-,v_{3}\rangle \end{pmatrix} =
\begin{pmatrix} M_{E} & 0_{3\times 3} \\ 0_{3\times 3} & M_{E} \end{pmatrix}
\begin{pmatrix} |+,1,1\rangle \\ |+,1,0\rangle \\ |+,1,-1\rangle \\  |-,1,1\rangle \\ |-,1,0\rangle \\ |-,1,-1\rangle \end{pmatrix}
\end{align}
with
\begin{equation}
M_{E}=\frac{1}{\sqrt{2}} \begin{pmatrix} 1 & 0 & 1 \\ -i & 0& i \\ 0 & \sqrt{2} & 0 \end{pmatrix}.
\end{equation}
We still have a block diagonal structure for the $L^{\uparrow}_{+}$ transformation 
\begin{equation}
U^{E}(\Lambda,0)=\begin{pmatrix} U^{E}_{R}(\Lambda,0) & 0 \\ 0& U^{E}_{L}(\Lambda,0) \end{pmatrix} = 
\begin{pmatrix} e^{-i \bm{J}^{E}\cdot \bm{\theta} + \bm{J}^{E}\cdot \bm{\phi}} & 0 \\ 
0&  e^{-i \bm{J}^{E}\cdot \bm{\theta} - \bm{J}^{E}\cdot \bm{\phi}}  \end{pmatrix} 
\end{equation}
with
\begin{equation}
(J^{E}_{i})_{jk}= -i \epsilon_{ijk},
\end{equation}
while the representation of parity remains invariant under this transformation. 
Next we go to the basis of well defined parity $ \{ |v_{i}, \pi  \rangle \}$ 
\begin{align}
\begin{pmatrix} |v_{1}, + \rangle \\ |v_{2}, + \rangle \\ |v_{3}, + \rangle \\  |v_{1}, - \rangle \\ |v_{2}, - \rangle \\ |v_{3}, - \rangle \end{pmatrix} 
= \frac{1}{\sqrt{2}} \begin{pmatrix} 1_{3\times 3} & 1_{3\times 3} \\ 1_{3\times 3} & -1_{3\times 3} \end{pmatrix}
\begin{pmatrix} |+,v_{1}\rangle \\ |+,v_{2} \rangle \\ |+,v_{3} \rangle \\  |-,v_{1}\rangle \\ |-,v_{2}\rangle \\ |-,v_{3}\rangle \end{pmatrix} .
\end{align}
In this basis parity operator is diagonal
\begin{equation}
\Pi^{\pi}(k)=\begin{pmatrix} \mathbbm{1}_{3\times 3} & 0_{3\times 3} \\ 0_{3\times 3} & - \mathbbm{1}_{3\times 3}  \end{pmatrix},
\end{equation}
while the generators take the form
\begin{equation}
\bm{J}^{\pi}= \begin{pmatrix} \bm{J}^{E} & 0 \\0 & \bm{J}^{E} \end{pmatrix} \qquad 
-i \bm{K}^{\pi}= \begin{pmatrix} 0&\bm{J}^{E}  \\ \bm{J}^{E} &0 \end{pmatrix} .
\end{equation}
The representation of the states in the basis is canonical in the same basis
\begin{align}
|v_{i}, +  \rangle \to \begin{pmatrix} \bm{e}_{i} \\ 0 \end{pmatrix}, \qquad 
|v_{i}, -  \rangle \to \begin{pmatrix} 0\\ \bm{e}_{i}  \end{pmatrix},
\end{align} 
such that a general state in this space can be written in terms the positive parity up component and  
negative parity down components  
\begin{equation}
\psi =\begin{pmatrix} i u \\  d \end{pmatrix}.
\end{equation}
Under and infinitesimal $L^{\uparrow}_{+}$ transformation we have
\begin{align}
 \delta u&= - i   (\bm{J}^{E}\cdot \bm{\theta}) u - i   (\bm{J}^{E} \cdot \bm{\phi})d
 =    \bm{\theta} \times \bm{u} + \bm{\phi}\times \bm{d} , \\
 \delta d &= - i   (\bm{J}^{E} \cdot \bm{\theta}) d + i  (\bm{J}^{E}\cdot \bm{\phi}) u
 =    \bm{\theta} \times \bm{d}  -  \bm{\phi}\times \bm{u} .
\end{align}
Define now $\psi^{ij}=\epsilon^{ijk}u^{k}$, $\psi^{0i}=d^{i}$ and $\theta^{l}=\frac{1}{2} \epsilon^{lrs}\theta_{rs}$ , $ \phi^{l}=\theta_{0l}$, these 
quantities transform as
\begin{align}
\delta \psi^{ij}&=  \frac{1}{2} \epsilon^{ijk}  \epsilon^{lrs}\theta_{rs} \psi^ {kl} +  \epsilon^{ijk} \epsilon^{klm}\theta_{0l}\psi^{0m} 
=- \frac{i}{2} (J^{\mu\nu}\theta_{\mu\nu})^{ij}_{~~ \gamma\delta}\psi^{\gamma\delta},\\
\delta \psi^{0i}&= \frac{1}{2} \epsilon^{ilm} \epsilon^{lrs}\theta_{rs}\psi^{0m} +  \frac{1}{2}  \epsilon^{ilm}\epsilon^{mjk}\theta_{0l} \psi^{jk} 
=- \frac{i}{2} (J^{\mu\nu}\theta_{\mu\nu})^{0i}_{~~ \gamma\delta}\psi^{\gamma\delta},
\end{align}
where
\begin{align}
(J^{\mu\nu})^{\alpha\beta}_{~~ \gamma\delta}= \frac{1}{2}\left( 
(J^{\mu\nu}_{V})^{\alpha}_{~\gamma} ~ g^{\beta}_{~\delta} + (J^{\mu\nu}_{V})^{\beta}_{~\delta} ~ g^{\alpha}_{~\gamma} 
- \gamma\leftrightarrow \delta \right) 
\end{align}
and $J_{V}^{\mu\nu}$ are the generators for the $(1/2,1/2)_{+-}$ irrep in Eq (\ref{JV}), from where we can see that the tensor 
$\psi^{\alpha\beta}$ transform in the antisymmetric product of two $(1/2,1/2)_{+-}$ representations
\begin{equation}
\psi^{\alpha\beta\prime}=\frac{1}{2}(\Lambda^{\alpha}_{~\gamma}\Lambda^{\beta}_{~\delta} 
- \Lambda^{\alpha}_{~\delta}\Lambda^{~\beta}_{\gamma})\psi^{\gamma\delta} 
\equiv  \Lambda^{\alpha\beta}_{~~\gamma\delta}\psi^{\gamma\delta} .
\end{equation}
In this basis, there is a pair of antisymmetric Lorentz indices $\alpha\beta$ for each spinor-like index $a$ thus the field is an antisymmetric 
rank-two tensor $\psi^{\alpha\beta}$ and the unit operator and the symmetric and $S^{\mu\nu}$ operator are given by 
\begin{align}
\mathbbm{1}_{\alpha\beta\gamma\delta}&= \frac{1}{2}(g_{\alpha\gamma}g_{\beta\delta} - g_{\alpha\delta}g_{\beta\gamma}),   \\
(S_{\mu\nu})_{\alpha\beta\gamma\delta}&= g_{\mu\nu} \mathbbm{1}_{\alpha\beta\gamma\delta}
- g_{\mu\gamma} \mathbbm{1}_{\alpha\beta\nu\delta} - g_{\mu\delta} \mathbbm{1}_{\alpha\beta\gamma\nu} 
- g_{\nu\gamma} \mathbbm{1}_{\alpha\beta\mu\delta}  - g_{\nu\delta} \mathbbm{1}_{\alpha\beta\gamma\mu}.
\end{align}
The operator tensors appearing in the projectors are
\begin{align}
(\Sigma_{\mu\nu})_{\alpha\beta\gamma\delta}&=\frac{1}{2}(g_{\mu\nu} \mathbbm{1}_{\alpha\beta\gamma\delta}
+(S_{\mu\nu})_{\alpha\beta\gamma\delta}) = g_{\mu\nu} \mathbbm{1}_{\alpha\beta\gamma\delta} 
- \frac{1}{2}( g_{\mu\gamma} \mathbbm{1}_{\alpha\beta\nu\delta} + g_{\mu\delta} \mathbbm{1}_{\alpha\beta\gamma\nu} 
+ g_{\nu\gamma} \mathbbm{1}_{\alpha\beta\mu\delta}  + g_{\nu\delta} \mathbbm{1}_{\alpha\beta\gamma\mu}), \\
(R_{\mu\nu})_{\alpha\beta\gamma\delta}&=\frac{1}{2}(g_{\mu\nu} \mathbbm{1}_{\alpha\beta\gamma\delta} 
-(S_{\mu\nu})_{\alpha\beta\gamma\delta}) = 
 \frac{1}{2}( g_{\mu\gamma} \mathbbm{1}_{\alpha\beta\nu\delta} + g_{\mu\delta} \mathbbm{1}_{\alpha\beta\gamma\nu} 
+ g_{\nu\gamma} \mathbbm{1}_{\alpha\beta\mu\delta}  + g_{\nu\delta} \mathbbm{1}_{\alpha\beta\gamma\mu}).
\label{Rtensor}
\end{align}
The Lagrangian in this basis reads 
\begin{align}
{\cal{L}}&= \partial^{\mu}\bar{\psi}^{\alpha\beta}(x)(\Sigma_{\mu\nu})_{\alpha\beta\gamma\delta} \partial^{\nu}\psi^{\gamma\delta}(x) 
- m^{2}\bar{\psi}^{\alpha\beta}(x) \psi_{\alpha\beta}(x) \nonumber \\
&= \partial^{\mu}\bar{\psi}^{\alpha\beta}\partial_{\mu}\psi_{\alpha\beta} - \partial^{\mu}\bar{\psi}^{\alpha\beta}\partial_{\alpha}\psi_{\mu\beta} 
 - \partial_{\mu}\bar{\psi}^{\mu\beta}\partial^{\nu}\psi_{\nu\beta} - m^{2} \bar{\psi}^{\alpha\beta}(x) \psi_{\alpha\beta} 
 \label{Ltmb}
\end{align}
The equation of motion for the field in this basis is
\begin{equation}
(\partial^{2}+m^{2})\psi_{\alpha\beta} - \partial_{\alpha}\psi_{\beta}+ \partial_{\beta}\psi_{\alpha}=0,
\label{eomtb}
\end{equation}
with 
\begin{equation}
\psi_{\alpha}=\partial^{\mu}\psi_{\mu\alpha}.
\end{equation}
Contracting Eq. (\ref{eomtb}) with $\partial_{\alpha}$ and using the result $\partial^{\alpha}\psi_{\alpha}=0$ due to the antisymmetry of 
the field, we get the constraint
\begin{equation}
\partial^{\alpha}\psi_{\alpha\beta}=0,
\label{coneomtb}
\end{equation}
and using this condition in Eq.(\ref{eomtb}) we obtain the Klein-Gordon equation for all the components of the field
\begin{equation}
(\partial^{2}+m^{2})\psi_{\alpha\beta}=0.
\end{equation}

As shown above, the equation of motion ensures that the negative parity component vanishes. In the Minkowskian basis 
for this space we get
\begin{equation}
(\partial^{2}\mathbbm{P}_{-})^{\alpha\beta\gamma\delta}\psi_{\gamma\delta}
=(R^{\mu\nu}\partial_{\mu}\partial_{\nu})^{\alpha\beta\gamma\delta}\psi_{\gamma\delta}
=\partial^{\alpha}\psi^{\beta} -\partial^{\beta}\psi^{\alpha}=0.
\end{equation}
It is worth to remark that this is a weaker condition than the $\psi_{\beta}=0$ condition in Eq.(\ref{coneomtb}). This will be important
in the massless case and it is quite similar to the structure of the $(1/2,1/2)_{+-}$ representation space where, in the massive case the 
vanishing of the unphysical parity component requires the field $V^{\alpha}$ to satisfy $\partial^{\alpha} (\partial\cdot V)=0$ but from 
the equation of motion we get the stronger Lorentz condition $\partial\cdot V=0$. In the massless case it is convenient to work in the 
Lorentz gauge but this does not completely fixes the gauge freedom which requires to use the corresponding parity projector.  Indeed, 
if $V_{\alpha}$ satisfies $\partial\cdot V=0$ so does $V^{\prime}_{\alpha}+\partial_{\alpha} f(x)$ whenever $f$ satisfies $\partial^{2}f=0$ 
which removes an additional degree of freedom leaving only two.

In the massless case, modulo a surface term, the Lagrangian 
\begin{align}
{\cal{L}}&=  \partial^{\mu}\bar{\psi}^{\alpha\beta}(x)(\Sigma_{\mu\nu})_{\alpha\beta\gamma\delta} \partial^{\nu}\psi^{\gamma\delta}(x)
\end{align}
is invariant under transformations in the unphysical (negative parity) component
\begin{equation}
\delta\psi_{\alpha\beta}(x)=  (R^{\mu\nu})_{\alpha\beta\gamma\delta}\partial_{\mu}\partial_{\nu}\zeta^{\gamma\delta} (x)
=  \partial_{\alpha}\zeta_{\beta} - \partial_{\beta}\zeta_{\alpha} 
\end{equation}
for arbitrary $\zeta^{\gamma\delta} (x)$, where $\zeta_{\alpha}=\partial^{\mu}\zeta_{\mu\alpha}$.  This transformation of course 
also a symmetry of the equation of motion
\begin{equation}
\partial^{2}\psi_{\alpha\beta} - \partial_{\alpha}\psi_{\beta}+ \partial_{\beta}\psi_{\alpha}=0,
\label{eomtbm0}
\end{equation}
but now the Lorentz-like condition $\psi_{\alpha}=0$ cannot be derived from this equation. If we impose this condition we get
a wave equation for the field $\psi_{\alpha\beta}$ but this does not completely fixes the gauge freedom. As shown 
above, the condition $\psi_{\beta}=0$ leaves only three independent components but if $\psi_{\alpha\beta}$ satisfy the $\psi_{\beta}=0$
condition, so does the field $\psi^{\prime}_{\alpha\beta}=\psi_{\alpha\beta}+ \partial_{\alpha}f_{\beta} - \partial_{\beta}f_{\alpha}$ 
whenever the vector field $f_{\alpha}$ satisfy the equation
\begin{equation}
\partial^{2} f_{\alpha}-\partial_{\alpha} (\partial\cdot f)=0.
\end{equation}
It is well known that a vector field satisfying this (Maxwell) equation has two independent degrees of freedom thus at the end 
the gauge symmetry impose five independent conditions on the six degrees of freedom of $\psi^{\alpha\beta}$ and we are left 
only with one propagating degree of freedom in the massless case. 

The tensor formulation for this representation has been considered in the study of diverse systems in 
particle physics and it would be important to explicitly establish the connection between our formalism based on first principles and 
those formulations. With this purpose, we can use
\begin{equation}
\partial^{\mu}\left( \bar{\psi}^{\alpha\beta}\partial_{\beta}\psi_{\mu\alpha} + \bar{\psi}_{\mu\alpha}\partial_{\beta}\psi^{\beta\alpha} \right)
=\partial^{\mu}\bar{\psi}^{\alpha\beta}\partial_{\beta}\psi_{\mu\alpha}+
\bar{\psi}^{\alpha}\psi_{\alpha}  ,
\end{equation}
to rewrite the Lagrangian in Eq. (\ref{Ltmb}) as
\begin{align}
{\cal{L}}&= \partial^{\mu}\bar{\psi}^{\alpha\beta}(\Sigma_{\mu\nu})_{\alpha\beta\gamma\delta} \partial^{\nu}\psi^{\gamma\delta}
- m^{2}\bar{\psi}^{\alpha\beta} \psi_{\alpha\beta} \nonumber \\
 &=\frac{1}{3}\psi^{\mu\alpha\beta} \psi_{\mu\alpha\beta} -  m^{2} \bar{\psi}^{\alpha\beta} \psi_{\alpha\beta} 
 - \partial^{\mu}\left( \bar{\psi}^{\alpha\beta}\partial_{\beta}\psi_{\mu\alpha} + \bar{\psi}_{\mu\alpha}\partial_{\beta}\psi^{\beta\alpha} \right)
\end{align}
where $\psi_{\mu\alpha\beta}$ is the totally antisymmetric tensor
\begin{equation}
\psi_{\mu\alpha\beta}= \partial_{\mu}\psi_{\alpha\beta} + \partial_{\alpha}\psi_{\beta\mu} + \partial_{\beta}\psi_{\mu\alpha}. 
\end{equation}
This tensor is invariant under the gauge transformations $\delta\psi^{\alpha\beta}=\partial^{\alpha}\zeta^{\beta} - \partial^{\beta}\zeta^{\alpha}$ 
for arbitrary vector field $\zeta^{\alpha}(x)$. In the massless case, modulo a surface term and an overall normalization our Lagrangian, 
coincide with the one used in the study of the classical description of the low energy interactions between open strings \cite{Kalb:1974yc}. 
The same Lagrangian has been previously used as a possibility for 
the low energy description of strongly interacting theories like QCD \cite{Chizhov:2011zz}. In the massive case it is equivalent to the formalism 
for the description of spin-one resonances used in the effective theory for QCD at low energies in the light sector \cite{Ecker:1988te}. 

The conclusion that there is only one propagating degree of freedom in the massless limit for the $(1,0)\oplus (0,1)$ representation 
space is certainly surprising and has been noticed long ago \cite{Kalb:1974yc}. Since we are posing here a first principles calculation 
which allows us to explicitly construct the one-particle states it would be interesting to scrutinize the kinematical origin of this result. 
This is easier in the spinor-like formalism and in the chiral representation where the spin and chiral structures 
are diagonal. In the chiral representation and in momentum space the state with well defined helicity $\lambda$ 
(eigenvalue of $h=\bm{J}\cdot \bm{n}$ ) has the following structure
\begin{equation}
\phi(\bm{p},\lambda)=\left(\begin{matrix} \phi_{R}(\bm{p},\lambda) \\  \phi_{L}(\bm{p},\lambda) \end{matrix} \right) .
\end{equation}
The equation of motion in Eq. (\ref{eomj1}) (projection onto well defined parity and Poincar\'{e} orbit subspace) impose the following conditions
\begin{align}
[1-\beta^{2}+ 2\beta h (\beta h+1)] \phi_{L}(\bm{p},\lambda) &=(1-\beta^{2}) \phi_{R}(\bm{p},\lambda), \\ 
[1-\beta^{2}+ 2\beta h (\beta h-1)] \phi_{R}(\bm{p},\lambda) &=(1-\beta^{2})\phi_{L}(\bm{p},\lambda) ,
\end{align}
where $\gamma=\cosh\phi=E/m$, $\beta=\tanh\phi=|\bm{p}|/E$, $h=\bm{J}\cdot\bm{n}$ and we used $h^{3}=h$. For $m\neq 0$ 
these equations are equivalent and we have only three independent conditions, leaving three of the six components of the spinor free, 
which correspond to the degrees of freedom of a spin-one particle. In the massless case however, $\beta\to 1$ and considering helicity 
eigenstates these equations reduce to
 \begin{align}
\lambda (\lambda +1)] \phi_{L}(\bm{p},\lambda) &=0, \\ 
\lambda(\lambda -1)] \phi_{R}(\bm{p},\lambda) &=0 .
\end{align}
We conclude that in the massless limit the right states can have only $\lambda=0,1$ and the left states can have only $\lambda=0,-1$. 
Since the construction of a parity eigenstates requires of  both left and right states of the same helicity we conclude that the only 
value of $\lambda$ which can be a parity eigenstate is the longitudinal mode with $\lambda=0$. Since parity is a good quantum 
number for a free particle, eigenstates of the free Hamiltonian (the propagating modes) must have well defined parity and for the 
$(1,0)\oplus (0,1)$ representation there is only one propagating mode. This is the opposite case to the $(1/2,1/2)_{+-}$ representation where 
in the massless case it is the longitudinal mode the one kinematically frozen in the massless limit and the propagating modes are the 
transverse modes $\lambda=\pm1$. Since in the Higgs mechanism for the $(1/2,1/2)_{+-}$ representation the longitudinal mode is taken 
from the scalar Higgs, it would be natural that the massless tensor field (the longitudinal mode of the $(1,0)\oplus (0,1)$ representation) 
take the transversal modes from a massless vector field in an alternative to the Higgs mechanism for this representation. This possibility 
has been put forth in \cite{Kalb:2009fj} and it would be desirable to study if a renormalizable theory can be constructed for the 
interactions of dark matter with $(1,0)\oplus (0,1)$ space-time structure using this mechanism. 


\section{Conclusions}

In this work, I consider the description of elementary massive particles from first principles. I start constructing the complete 
set of commuting operators for a free relativistic massive particle. In general, for a 
quantum system a choice for this set is given by the Casimir operators of its symmetry group, the generators of the Cartan 
subalgebra of this group (the subset of commuting generators of the subgroup of continuous symmetries connected to the 
identity) and the discrete hermitian symmetry operators commuting among them and with the previous operators. 

At the classical level, a free relativistic particle has the same symmetries as the Minkowski space-time and 
the symmetry group is the set of transformations leaving invariant the length element $ds^{2}=dx^{\mu}dx_{\mu}$, 
named Poincar\'{e} group. We review the well known structure of this group.  

At the quantum level, Poincar\'{e} transformations are implemented by unitary operators $U(\Lambda,a)$ in the space of 
quantum states. Using the group composition rule we show that a suitable complete set of commuting observables for a 
quantum relativistic massive particle is given by $\{ H, P^{2}, W^{2}, J_{3}, \Pi \}$, 
where $\Pi=U({\cal{P}},0)$ is the quantum operator for parity. Free particle 
quantum states can be labelled by the eigenvalues of these operators (the good quantum numbers), 
$| E, m^{2}, -m^{2}j(j+1), \lambda, \pi\rangle$. These states, named Poincar\'{e} states here, are not eigenstates of 
$\bm{P}$, only of $| \bm{P}|^{2}$ with eigenvalue  $|\bm{p}|^{2}=E^{2}-m^{2}$. The only frame where these states have well 
defined $\bm{P}$ quantum numbers is the rest frame where the eigenvalues of $H$ and $P^{2}$ fix univocally the eigenvalues 
of $P^{\mu}$ to $k^{\mu}=(m,0,0,0)$. Rest frame Poincar\'{e} states can be labelled also by the eigenvalues of $P^{\mu}$, 
and we use the notation $| m, m^{2}, -m^{2}j(j+1), \lambda, \pi\rangle \equiv | k, m^{2}, j, \lambda, \pi \rangle $.  

Eigenstates of well defined momentum can be constructed acting with the boost operator $U(L(p),0)$ on the rest frame 
Poincar\'{e} states where $p^{\mu}=(L(p))^{\mu}_{~\nu}k^{\nu}$. These states, denoted as 
$| p, m^{2}, j\rangle_{\lambda\pi }$, satisfy non-trivial kinematical conditions stemming from the parity
eigenvalue equation satisfied by the rest frame Poincar\'{e} states. Using the general factorization of Poincar\'{e} transformations 
into the product of a space-time translation and a homogeneous Lorentz transformation I solve the $L^{\uparrow}_{+}$ algebra, 
construct matrix representations for operators and states and study these kinematical conditions for the $(j,0)\oplus (0,j)$ 
and $(1/2,1/2)$ quantum representations. 

For $j=1/2$, the kinematical constraint due to parity yields Dirac equation. The Dirac gamma matrices in this perspective are the 
covariant companions of parity which in the rest frame is $\Pi (k)=\gamma^{0}$ and the Dirac algebra has a kinematical origin.

For $j=1$, the kinematical condition 
yields a second order equation of motion for the $(1,0)\oplus (0,1)$ representation in a spinor-like formalism, where the 
$| p, m^{2}, j\rangle_{\lambda\pi }$ states are represented by "spinors" with six components. This field satisfy kinematical constraints 
related to the vanishing of the undesired (unphysical) parity component contained in the $(1,0)\oplus (0,1)$ space. At the Lagrangian 
level these are second class constraints which can be solved and it has been shown recently that a sensible quantum field theory 
is obtained. The structure of the theory makes it appealing for dark matter and the corresponding effective theory for interactions 
with standard model fields and its phenomenology has been recently worked out 
\cite{Gomez-Avila:2013qaa,Napsuciale:2015kua,Hernandez-Arellano:2018sen, Hernandez-Arellano:2019qgd}.
 
I trace the parity quantum numbers in the $(1/2,1/2)$ spaces 
generated from the $(1/2,0)\otimes (0,1/2)$ and $(0,1/2)\otimes (1/2,0)$ tensor products and show that parity requires to work with 
the $[(1/2,0)\otimes (0,1/2)] \oplus [(0,1/2)\otimes (1/2,0)]$ representation which can be decomposed into the direct sum of two 
$(1/2,1/2)$ subspaces of well defined parity: $(1/2,1/2)_{+-} \oplus (1/2,1/2)_{-+}$ where the subindex denote the parity of the 
$j=0$ and $j=1$ Poincar\'{e} content of these representations. We construct the rest frame Poincar\'{e} states in the $(1/2,1/2)_{+-}$ 
representation space and show that the kinematical constraint due to parity and the $P^{2}$ eigenvalue equation yields a second 
order spinor-like equation of motion for the states with a constraint which eliminates the unphysical parity component. 

A suitable Lagrangian for the derivation of this equation is constructed and the Lagrangian has a soft $m^{2}\to 0$ limit. In the massless 
limit, the constraint eliminating the unphysical parity component is lost and the kinetic operator is singular, but a symmetry 
emerges related to arbitrary changes in the unphysical parity component. Introducing a parity fixing term in the Lagrangian with 
a Lagrange multiplier well defined parity is recovered and the kinetic operator is invertible but depends on the value of the Lagrange 
multiplier. 

I show that in a specific basis for the states in $(1/2,1/2)_{+-}$ the spinor-like components transform as the components of a Lorentz 
four-vector, the spinor-like equation of motion stemming from parity and $P^{2}$ in the massive case is the Proca equation and the 
constraint eliminating the unphysical parity component is the Lorentz condition. In the massless limit, the emergent symmetry related to 
changes in the unphysical parity component is the celebrated gauge symmetry of the massless vector field and the 
parity-fixing term in the Lagrangian is the conventional gauge-fixing term, which reveals parity as the space-time origin of gauge 
symmetry.  

A similar structure is obtained in the massless limit of the formalism for the $(1,0)\oplus (0,1)$ representation space. In this limit the 
constraint eliminating the unphysical parity is lost and the kinetic operator becomes singular but a symmetry related to arbitrary changes 
in the unphysical parity sector emerges. Introducing a parity-fixing term in the Lagrangian with a Lagrange multiplier we recover well 
defined parity and an invertible kinetic operator. The propagator has a good ultraviolet behavior but depends on the Lagrange 
multiplier.

In a specific basis for states in the $(1,0)\oplus (0,1)$ representation space we show that every spinor-like index transform as a pair of 
anti-symmetric Lorentz indices. In this basis, the equation of motion in the massive case yields the Kalb-Ramond equation 
\cite{Kalb:1974yc} and the condition for the vanishing of the unphysical parity component is the analogous of the Lorentz condition 
for the Kalb-Ramond field.  In the massless case the emergent symmetry in this basis is the gauge symmetry of the massless 
Kalb-Ramond field  and the parity-fixing term corresponds to a gauge fixing term.  The propagator has a good ultraviolet behavior but depends 
on the value of the Lagrange multiplier, i.e, is gauge dependent. The gauge symmetry leaves only one degree of freedom in the 
massless case. We study this result on the light of the explicit calculation of the states solving the $L^{\uparrow}_{+}$ algebra 
concluding that the transversal modes of the spin-one system described by this equation are kinematically frozen in the massless 
limit due to parity and only the longitudinal mode can be a parity eigenstate and survives as a propagating mode. 

All these results neatly establish parity as the space-time symmetry behind the equations of motion for spinning particles and 
as the space-time origin of gauge symmetry. 

\section{acknowledgements}
The content of this paper is an evolution of work done previously with my collaborators, mainly Mariana Kirchbach, Sim\'{o}n Rodr\'{i}guez
and Selim G\'{o}mez-\'{Avila}. I thank them all for fruitful discussions on the subject.

\bibliographystyle{prsty}
\bibliography{spacetime}

\end{document}